\documentclass{article}
\usepackage{amsmath}
\usepackage{amssymb}
\usepackage{makeidx}
\usepackage{latexsym}
\usepackage[dvips]{graphicx}
\usepackage{textcomp}

\begin{document}

\title{{\huge Piezonuclear reactions }}
\author{{\large Fabio Cardone}$^{1,2}${\large , Roberto Mignani}$^{2-3}$%
{\large \ and Andrea Petrucci}$^{1}$ \\
$^{1}$Istituto per lo Studio dei Materiali Nanostrutturati (ISMN -- CNR)\\
Via dei Taurini - 00185 Roma, Italy\\
$^{2}$GNFM, Istituto Nazionale di Alta Matematica "F.Severi"\\
\ Citt\`{a} Universitaria, P.le A.Moro 2 - 00185 Roma, Italy\\
$^{3}$Dipartimento di Fisica \textquotedblright
E.Amaldi\textquotedblright ,
Universit\`{a} degli Studi \textquotedblright Roma Tre\textquotedblright \\
\ Via della Vasca Navale, 84 - 00146 Roma, Italy\\
\\
} \maketitle
\date{}

\begin{abstract}
In this paper, we deal with the subject of piezonuclear reactions,
namely nuclear reactions (of new type) triggered by pressure waves.
We discuss the experimental evidences obtained in the last two
decades, which can be summarized essentially as follows: experiments
in cavitation of liquids, where transmutation of elements, creation
of elements and emission of neutrons have been observed; emission of
neutrons in brittle failure of solids subjected to mechanical
pressure; alteration \ of the lifetime of un unstable element
(thorium) subjected to cavitation. A theoretical model to explain
these facts is proposed. Future perspectives of these experimental
and theoretical investigations are also underlined.
\end{abstract}

\newpage

\section{Introduction}

The application of elastic waves of suitable power and frequency (in
particular ultrasounds) to a liquid with gas dispersed in it gives
rise to the phenomenon known as acoustic cavitation~\cite{cg,bren}.
The main physical effects occurring in a cavitated liquid can be
accounted for in terms of a hydrodynamic model based on the
formation and the collapse of imploding gas bubbles in the
liquid~\cite{cg,bren}. The processes occurring during the collapse,
and the collapse itself, are quite complex and a good deal of
research has been carried out in order to clarify their physical and
chemical aspects.

In particular, theoretical and experimental studies on
sonoluminescence (i.e. the emission of light bursts from a cavitated
liquid \cite{RMP}) led to conclude that the temperature in the core
of a collapsing bubble can
reach $10^{6}$ $%
{{}^\circ}%
K$ and more \cite{flan,chen}. Actually, even simple-minded
calculations show that cavitation can produce an extreme
concentration of energy for unit time in the collapsing gaseous
bubble. To this regard, let us consider a typical
bubble, whose initial radius, as experimentally found, is of the order of 10$%
^{-6}$ m. If we suppose that nothing can stop the contraction down
to a radius of atomic dimensions, namely 10$^{-10}$ m, then, by
assuming that the total initial power stored within the bubble be
constant, the surface power density at the end of contraction will
be 10$^{8}$ times greater than the initial one. For an initial power
of 100 $W$, this means that within the
final bubble we will have an equivalent temperature of 10$^{11}$ $%
{{}^\circ}%
K$ (corresponding to an energy for particle of 10 $MeV$). So high
temperatures could cause thermonuclear fusion (\emph{sonofusion or
acoustic inertial confinement fusion, AICF}).

The research on AICF (pioneered by Flynn in 1982~\cite{crum})
carried out in the last years \cite{ar1}-\cite{kriv} attempted to
produce known nuclear reactions by means of pressure or ultrasounds
and cavitation. To this aim, the solutions involved contained
deuterium and/or unstable nuclides. Let us notice that this is by no
means a completely new approach to standard nuclear reactions
(except for the use of cavitation). Indeed, we recall that in the
past some investigations~\cite{dieb}-\cite{wint} have highlighted
the ability of pressure and shock waves to generate autocatalytic
fission-fusion reactions in compounds containing also uranium,
tritium and deuterium.

On the other hand, experiments on the electric explosion of titanium
foils in liquids \cite{uru1}-\cite{uru3} have evidenced changes in
the concentrations of elements and alteration of the secular
equilibrium of unstable nuclides. Since one might argue that the
shock waves caused by the foil explosion act on the matter in a way
similar to ultrasounds in cavitation, all such experiments, together
with those obtained by cavitating liquids, support the evidence for
nuclear reactions induced by high mechanical pressures
(\emph{piezonuclear reactions, }piezo from the ancient Greek verb
\emph{piez\'{e}in} which means to press).

Moreover, since cavitation in liquids is basically a mean to
generate pressure waves acting on the elements diluted in solutions,
we hypothesized that \emph{mechanical waves may give rise to
piezonuclear effects also in solids subjected to pressure during
crashing failure}. In other words, compenetration of molecular
layers in solids would be the analogous of cavitation in liquids.
This hypothesis was tested and proved indeed to be true \cite{carp3,
carcarp}.

These considerations stress the need for (and the importance of)
experiments looking at the physical features and implications of
piezonuclear reactions, with special emphasis on the mass spectrum
of the samples (either liquid or solid) before and after cavitation
and on the possible emission of ionizing radiation and neutrons.

To this aim, in the last years, we carried out a series of
experiments on piezonuclear reactions \cite{carmig1}-\cite{carp}.
They can be roughly
divided in three groups: 1) experiments on cavitating water \cite{carmig1}-%
\cite{carmig4}; 2) experiments on cavitating salt solutions \cite{carcher1}-%
\cite{carmig5}; 3) experiments on fracture of solids \cite{carp,
carcarp}. In the first group, we looked for transformations of
elements, whereas the other two sets of experiments were devoted to
the detection of emitted neutrons. Two points need to be stressed.
First, our research, although it might seem to deal with the same
physical terms such as nuclear reactions, nuclear radiation and to
point towards the same technological direction, is based on a
different theoretical framework \cite{carmig6}-\cite{carmig8},
according to which nuclear processes (not necessarily of the
standard type) can be induced in stable nuclides too only by
suitable mechanical waves, i.e. without any application of
radioactive or nuclear active substances. Then, it never involved
any radioactive source or unstable nuclide (except
for the case of \cite{carmig5}), unlike other experiments~\cite{tal1}-\cite%
{kriv},\cite{uru1}-\cite{uru3}. Our work concentrated on absolutely
stable
elements\footnote{%
These stable elements are all but deuterium, since we are not
looking for Deuterium-Deuterium fusion.}, and in this sense it is
moving along a parallel path with respect to the other research
streams.

The paper is organized as follows. In Sect.2, we discuss all the
experiments carried out by cavitating water. The results obtained by
cavitation of liquid solutions of metal salts are described in
Sect.3, where the fundamental role of iron salts in order to get
neutron emission is stressed. Sect.4 deals with the problem of the
effect of cavitation on a solution containing thorium, which
apparently shows its reduction at a rate 10$^{4}$ times faster than
its natural decay. In Sect.5 we discuss the evidence for
piezonuclear reactions in brittle fracture of solids, in which
neutron emissions is observed too. Sect.6 is devoted \ to a brief
comparison with other experiments of similar kind (although a deep
discussion with LENR phenomenology deserves a paper by its own). In
Sect.7 we propose a possible theoretical explanation of piezonuclear
reactions, partly classical and partly based on the formalism of a
deformed Minkowski space-time (in which the metric coefficients are
functions of energy). The last section contains some possible
developments and perspectives.

\section{Piezonuclear Reactions in Cavitated Water}

In our first three experiments on cavitation
\cite{carmig1}-\cite{carmig4}, we restricted ourselves to detect the
mass spectrum changes, after cavitation, together with possible
presence of ionizing radiation. As regards the substance undergoing
cavitation, two main considerations led us to choose the deionized
and distilled water. In first place, we know from both theoretical
and experimental studies that its phenomenological behaviour during
cavitation satisfies the Rayleigh-Plesset equation. In this respect,
we remark that, although at the present time we don't have a simple
theory of phenomena (such as sonoluminescence), connected to the
cavitation process, and the use of this equation was not accepted by
all researchers \cite{RMP}, the latter is, however, the starting
point for a determination of the macroscopic phenomenological
parameters needed to implement the cavitation itself.

\subsection{First Experiment}

In the first experimental work (carried out at Perugia University in
1998)
\cite{carmig1,carmig2}, we utilized a new type of\textbf{\ }sonotrode\textbf{%
\ }(\textquotedblright cavitator\textquotedblright ) with a very
long
working time ($>$ 30 $min$)\footnote{%
This new kind of cavitator was developed by P. Diodati at Perugia University.%
}. Its main feature consisted in the cooling system, that allows one
to keep constant the resonance frequency for a time interval much
longer that possible with the presently available sonotrodes. This
cavitator was composed by a titanium tip shaped with an exponential
profile, connected to a column of piezoelectrics cooled by oil in a
pressure circuit with dry ice choke.

Through such a device we subjected to cavitation a sample of
bidistilled and deionised water at room temperature. The vessel of
the cavitator was made of optical flint glass. During the cavitation
process, the titanium tip of the cavitator was directly immersed and
the water surface was free. The water sample was subjected to
ultrasounds without stopping for a total time of 210 $min$ at the
constant power of 630 $W$ and at the frequency of 20 $kHz$, in
order to induce cavitation according to the Rayleigh-Plesset equation~\cite%
{bren}.

We performed a gamma spectrometry on samples of cavitated and
uncavitated water without finding any radioactivity increase in
cavitated water. After cavitation, we analyzed the cavitated water
sample, confining our analysis to the stable chemical elements (from
Z = 1 to Z = 92), and we compared the new results with those
obtained from the uncavitated water. The analysis of the water both
before and after cavitation was carried out by three different
procedures, reaching the precision of 1 $ppb$ and with a standard
deviation on concentration given by $\sigma $= 10$^{-5}$ $%
\mu
g/l$. The three procedures consisted in:

- mass atomic absorption (ICP);

- cyclotron spectrometry (ICR);

- mass spectrometry (MS).

All these measurements have been carried out using different
apparatuses and by different personnel staff.

Therefore, the concentration of each element was measured at least
three times. More precisely: the ICP and the ICR measurements were
carried out for all elements; as a check, the ICP measurement was
repeated for the elements with decreasing and increasing
concentrations; the MS measurements were carried out only for the
elements from H to Pd.

We carried out also a measurement of device background, without
water, for each vacuum chamber used.

Very surprisingly, we found relevant changes in the concentrations
of the
elements (in units of $%
\mu
g/l$) in the cavitated sample (despite the very low original
concentration). In order to asset the changes in the chemical
elements, the variation factors have been accepted only for a value
$\geq $2 in the concentration ratios in water after and before
cavitation. Evidence was found for 10 increasing and 19 decreasing
elements. Let us notice, in particular, that the decrease concerned
stable elements with low mass number.

We checked the possible contributions to changes due to impurities,
possibly arising from the titanium tip of the cavitator and the
flint glass of the vessel, by three different methods:

- mass atomic absorption (ICP) on cavitated water sample;

- electron microscopy on dusts of tip and vessel;

- X-ray microanalysis, carried out by two different laboratories on
both the dusts of tip and vessel and the dry residues of the two
(cavitated and uncavitated) water samples.

The obtained results let us exclude contributions due to impurities
to the observed concentration changes \cite{carmig1,carmig2}.

During cavitation, we looked for the possible emission of radiation
by putting, on the external walls of the vessel, slabs of colloid
LR115, sensitive to ionisation energies in the range 100 $KeV$- 4
$MeV$, which are typical e.g. of $\alpha $-particles. We measured,
by means of the same kind of slabs, the background radiation in the
laboratory room where water cavitation was being carried out, and we
found a flux intensity of (210 - 150) $Bq$ $m^{-2}$. We analyzed the
slabs and observed no significant differences between the two sets
of slabs, those put on the vessel walls and those which recorded the
background radiation. Subsequently, five months later, we carried
out again a $\gamma $-spectrometry on samples of both uncavitated
and cavitated water without finding any appreciable difference, thus
confirming the previous evidence.

A basic point to be stressed is that, even if the cavitation gave
rise to an increase of the concentration of some elements and to a
decrease of others, \textit{the number of protons was actually
conserved. }This is easily seen
if we take into consideration the difference $\Delta $, defined as%
\begin{equation}
\Delta =\sum_{i}R_{i}Z_{i}-\sum_{d}R_{d}Z_{d}
\end{equation}%
where $R_{i}$ ($R_{d}$) is the average concentration ratio after
cavitation of the i-th (d-th) increasing (decreasing) element, and
$Z_{i}$ ($Z_{d}$)
the corresponding atomic number. From our data we found:%
\begin{equation}
\Delta =3.2
\end{equation}%
which is actually negligible, being about 10$^{-3}$ of the number of
protons
involved (on account of the variation factors), say $R_{i}Z_{i}$ , $%
R_{d}Z_{d}$ for each element undergoing a concentration change. On
the contrary, \emph{the number of neutrons between increasing and
decreasing elements was not conserved. }

Moreover, a huge increase in the concentration of uranium was found,
of about two orders of magnitude. This result led us to perform the
second experiment.

\subsection{$\ $Second Experiment:Transuranic Evidences}

The measurements of the first experiment were confined to the stable
chemical elements. We\ therefore performed a second cavitation
experiment and analyzed the mass composition of cavitated water by a
spectrometer in the mass region 210$<$M$<$271 \cite{carmig3}.

We subjected to cavitation by a standard sonotrode a mass of
bidistilled and deionised water of about 30 $g$, contained in a
pyrex vessel at room temperature. The cavitation was carried out in
an underground laboratory at the Tuscia University of Viterbo.

Four mass measurements were carried out on samples from water
cavitated one, two, three and four times. Therefore, the whole
cavitation time ranged from 10 $min$ for the first water sample to
40 $min$ for the last water sample. Each measurement was performed
immediately after each cavitation run. We confined ourselves to
merely counting the different masses identified by the spectrometer.
An error $\Delta M=0.1$ was accepted in the mass value
determination.

\begin{figure}
\begin{center} \
\includegraphics[width=0.8\textwidth]{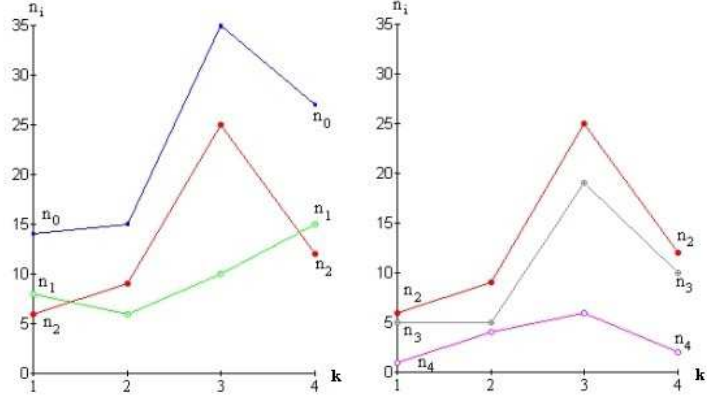} \caption{\textit{Numbers n$_i$ ( i = 0, .., 4) of nuclides vs. the number k ( k = 1, .., 4) of
cavitation runs. Left: Behavior of n$_0$ , n$_1$ and n$_2$; Right:
behavior of n$_2$ , n$_3$ and n$_4$.}}\label{unst}
\end{center}
\end{figure}

Let us label by $k$ and $i$ ($k=1,..,4$; $i=0,..,4$), respectively,
the four measurement runs carried out and the mass intervals (the
total interval\ 0: 210$<$M$<$271, was split in the two subintervals
1: 210$<$M$<$238, and 2: 238$<$M$<$271; the subinterval 2 was in
turn split in the subintervals 3: 238$<$M$<$264, and 4:
264$<$M$<$271). The numbers $n_{ik}$ are defined as
\begin{equation}
n_{ik}\equiv N_{ik}-N_{Bi},
\end{equation}
where $N_{ik}$ is the number of nuclei measured in the $i$-th mass
interval in the $k$-th water sample cavitated $k$ times, and
$N_{Bi}$ is the number of nuclei measured in the $i$-th mass
interval in the background/blank measurement (which of course
includes the masses of the calibration compound).

In Fig.\ref{unst} the numbers of nuclides $n_{i}$ ($i=0,..,4$) are
plotted as functions of the cavitation runs $k$. On the left it is
seen that the excess in the number of masses identified in run 3, in
the water sample cavitated 3 times, has to be ascribed to masses in
region 2, 238$<$M$<$271. In such a region 2, the contribution to the
excess in the number of masses appears to be due to masses in the
transuranic region,
238$<$M$<$264, as shown on the right in Fig.\ref%
{unst}.

One might argue that the excess in the number of masses could be due
to the formation of macromolecules by processes like \textit{e.g.
}dimerization. However, such a possibility is unable to explain why
the excess is just concentrated in a definite mass interval,
\textit{i.e.} in the transuranic region. On the other hand, possible
contributions from cosmic-rays induced phenomena (like spallation)
even at altitudes less than 200 $m$ a.s.l.--- as it is indeed the
case of the experiment sites --- could give rise to an excess of
light nuclei, but cannot explain an increment of heavy nuclei (even
less the presence of nuclei heavier than uranium). Moreover, the
actual presence of transuranic elements is supported by the huge
increase of uranium found in the first experiment, which might just
be explained in terms of the formation and subsequent decay of
transuranic elements. The results of the first experiment do agree
with the results of the second one, showing the rearrangement of
elements, with the decrease in the concentrations of light-medium
elements, and the increase in those of the heavy ones. Although a
definite conclusion can only be drawn by carrying out a further
experiment in the whole mass range, including both stable and
transuranic elements, it is reasonable to assume that the excess of
masses in the transuranic region are compensated by a simultaneous
deficit of masses in the stable element region.

Such a picture is supported by the different behavior of the $n_{i}$'s ($%
i=1,2,3,4$). The net increase of $n_{1}$ is due to the fact that in
the mass range 210$<$M$<$238\ there are mainly stable nuclei, and
therefore their number increases following the production. On the
contrary, in the range $Z=92\div 114$ and 238$<$M$<$270, there are
only unstable radionuclides (either experimentally known or
theoretically predicted). Thus, their number first increases, then
decreases (after the third cavitation) as soon as the decay rate
overcomes the rate of production.

\subsection{Third Experiment:Europium Isotope Production}

We have seen that the first two experiments provided evidence for a
change in concentrations of stable and transuranic chemical elements
in cavitated water \cite{carmig1,carmig2}. The third experiment was
aimed at looking for the production of the so-called rare earth
elements, and was performed at the University of Rome
\textquotedblright La Sapienza\textquotedblright\ \cite{carmig4}. To
cavitate water, the same sonotrode device of the second experiment
was employed, but with a reduced power setup (still able to induce
cavitation), in order to account for the different contributions to
the phenomenon due to both the cavitation and the ultrasounds at
different powers. The sonotrode tip, shaped like a truncated cone,
was plunged into an open vessel at atmospheric pressure, filled with
water at room temperature. The water surface was free. The sonotrode
was cooled by air at room
temperature (20$%
{{}^\circ}%
C$) and had a working frequency $\nu $ = 20 $kHz$, and transmitted
power $P$
= 100 $W$. The continuous operation folding time of this sonotrode was 15 $%
min$, followed by a cooling period (15 $min$) of the column of
piezoelectrics. During the cooling period the sonotrode was off.

It was chosen to examine, through ICP, two mass intervals, from 90 to 150 $%
amu$ and from 200 to 255 $amu$, since they include also the rare
earth elements we were looking for.

For every mass value in the two intervals, subdivided into steps of 0.01$amu$%
, the results of the ICP mass count were analyzed. The count series
coincided with the measure series; the latter were obtained using
the count data acquisition program PQ Vision supplied by Thermo
Elemental. The aim of this analysis was to highlight the count
variations (decrease or increase compared with the previous count).

For each mass the upper limit for the count background was
evaluated, including both blank and noise. One took into
consideration the masses whose upper count was higher than this
upper limit. A given mass was taken into account only if its count
was at least twice greater than the upper limit.

This criterion was applied to the counts obtained, for the same mass
values, from measures performed with scanning times of both 10 $s$
and 150 $s$.

No mass whose counts satisfied the described criterion was
identified for scanning times of 150 $s$, whereas only one mass was
found for the scanning time of 10 $s$, namely $M$ =(137.93 $\pm$
0.01) $amu$.

\begin{figure}
\begin{center} \
\includegraphics[width=0.8\textwidth]{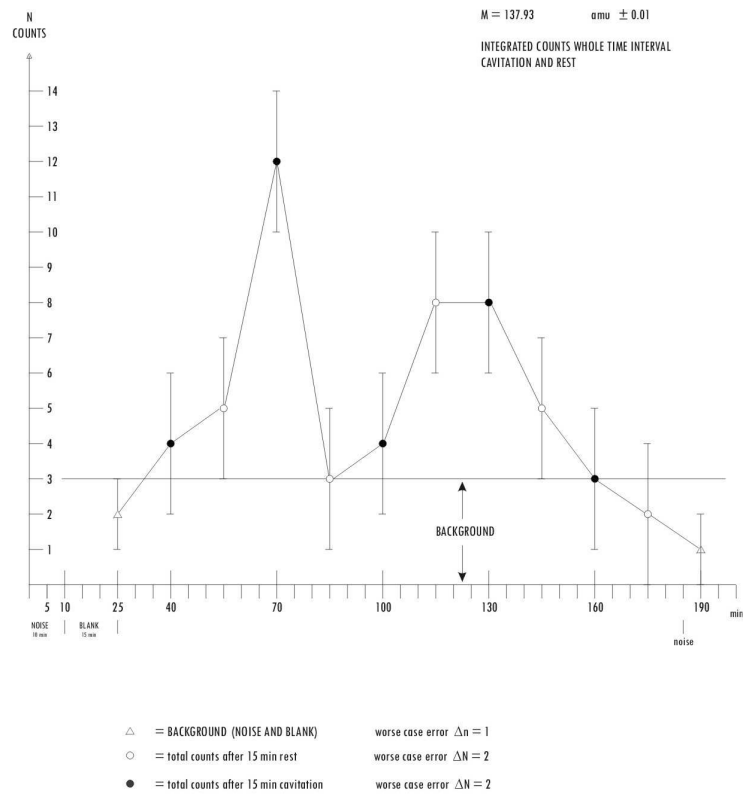} \caption{\textit{Integrated counts for M
= 137.93} $\mathit{amu}$ \textit{\ in the third experiment of
cavitating water}.}\label{eur}
\end{center}
\end{figure}

We performed differential measurements with a time interval of 300
$s$, corresponding to the cavitation interval time. Then one got the
differential counts, from which we derived the integral counts
plotted in Fig. \ref{eur} as function of the cavitation interval
time. In this way the data have been ensured from instrumental
pile-up effects.

The identification of the observed peak with a given radionuclide
required
the determination of the lifetime. This could be done by analyzing Fig.\ref%
{eur}, and interpreting it as two subsequent cycles of production
and decay of the observed element\cite{carmig3,carmig4}. Thus it was
possible to evaluate\ its halftime $T_{1/2}=$(12$\pm $1) $s$ within
1.5 $\sigma $.

From the tables of nuclides \cite{chart} we got as possible
candidate the isotope of europium $Eu_{63}^{138}$
\cite{carmig3,carmig4}.

During the first experiment the concentration of $Eu_{63}$ was not
changed. It is known that the abundance on Earth of stable europium
is less than 1.06 $ppm$ (the natural $Eu$ is a mixture of two
isotopes, $Eu_{63}^{151}$ with a percentage abundance of 47.77\% and
$Eu_{63}^{153}$ with a percentage abundance of 52.23\%). The
candidate identified during the third experiment does not exist in
nature; it is an artificial radionuclide (discovered only in 1995-97
\cite{chart}) that can be produced at the present time in nuclear
reactors and by synchrotrons.

There are two ways whereby $Eu_{63}^{138}$ can be produced: by
nuclear fission or by nuclear fusion. The former process requires
less energy.
However, from the results of the first two experiments \cite{carmig1,carmig2}%
, the quantity of heavy nuclei which can produce $Eu^{138}$ by
nuclear fission is very much smaller (by two-three orders of
magnitude) than that of the intermediate nuclei which can produce it
by fusion (on account also of the impurities of the tip). As a
matter of fact, a rough estimate (just based on the detected
abundance of heavy nuclei in the second experiment) yields a
probability of 10$^{-6}$-10$^{-8}$ for the production of $Eu^{138}$
by fission. Moreover, nuclear fusion is the only possible
explanation of the changes in concentration of stable elements,
induced by cavitation, observed in the first experiment.

\bigskip

\section{Piezonuclear Reactions in Cavitated Solutions of Stable Elements
and Evidence for Neutron Emission}

We recall that, in our first experiment (see Sect.2), proton number
was practically conserved, whereas neutron number was apparently
not. This constitutes an indirect hint of some sort of neutron
production in such cavitation processes. Since, as is well known,
nuclear reactions in most cases involve neutron emission, it is a
fundamental issue to check whether neutrons are produced indeed in
processes possibly involving piezonuclear
reactions. We point out that e.g. the experiments of refs.\cite{tal1}-\cite%
{tal3} have shown that cavitation of deuterated acetone can produce
neutrons. In order to shed some light on this issue of neutron
emission during cavitation, in 2004-2007 we carried out some
experiments in which we cavitated controlled solutions of salts in
water at CNR National Laboratories (Rome 1 Area) and Italian Armed
Forces technical facilities (also with the support of ENEA, Ansaldo
Nucleare and ARPA-Lazio). We focused our attention on ionizing
radiation and neutron emission. The details of these experiments are
reported in the following.

\subsection{Experimental Equipment}

The employed ultrasonic equipment was the robust ultrasound welder
DN20/2000MD by Sonotronic~\cite{sono}. We slightly modified the
piezoelectrics and the sonotrode configuration in order to provide
the equipment with a compressed air cooling system which allowed it
to work for 90 minutes without stopping, at a frequency of 20 $kHz$.
As cavitation chamber, we used a Schott Duran\circledR ~vessel made
of borosilicate glass of 250 $ml$ and 500 $ml$~\cite{duran}. The
truncated conical sonotrode that conveyed ultrasounds was made of
AISI grade 304 steel. In all the experiments, the cavitated
solutions were made using deionized and bidistilled water (18.2
$M\Omega $).

A magnetometer was used in order to take under control the local
magnetic field (always found compatible with the local magnetic
field of Earth, measured in absence of cavitation) and along with it
possible currents generated by the converting piezoelectric units
that might have affected the electronics of the geiger counters and
of the gamma spectrometer.

Let's now focus our attention on the technique used to reveal the
possible neutron emission that we may expect during cavitation from
the results of our previous experiments (see Sect.2). The only hint
that we got from them is the non-conservation of the number of
neutrons (according to the mass spectrometer analyses), which
suggests a possible neutron emission, but does not say anything
about their spectrum, their isotropy and homogeneity in space and
their constancy in time. This wide range of possibilities convinced
us that the first step to be moved in order to ascertain this hint
was just to reveal the presence of neutrons in a sort of a 'yes or
no' detecting procedure, by leaving a complete and more exhausting
proper measurement to a second higher and more accurate level of
investigation, grounded on the possible positive answer from this
first level of inquiry. Thus, we decided at first to use neutron
passive detectors which are capable of integrating neutron radiation
within their energy range, regardless of the time feature of their
emission. Such devices do not require any sort of adaptive
electronic calibration which would be necessary with an electronic
detector in order to track and follow an emission that, as far as we
know, could be the most variable one in terms of energy and time.
The passive detectors that we utilized were the so-called Defenders
and the CR-39 plates. Once that an initial but solid evidence of
neutron emission was gathered by these passive detectors, we
increased the quality of our investigation by moving to electronic
Boron Trifluoride detectors. Before presenting the experiments and
their results, it is important to stress at this stage some features
of the detectors and state what was done in order to keep them under
control.

\begin{figure}
\begin{center} \
\includegraphics[width=0.8\textwidth]{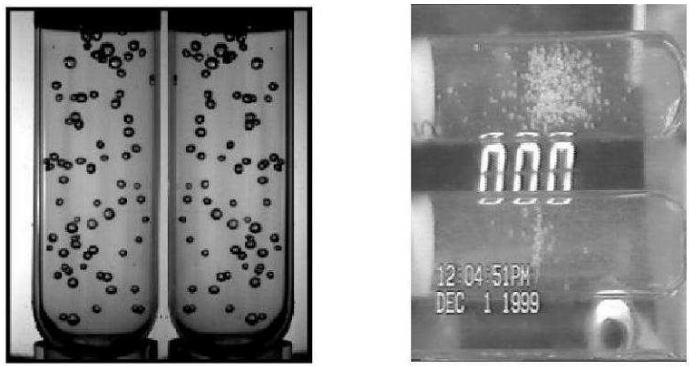} \caption{\textit{Morphology and
distribution of bubbles produced in a Defender by the passage of
neutrons (left); heavy ions (right). In the second picture, the ion
beam goes from bottom to top.}}\label{neutr-ion}
\end{center}
\end{figure}

The Defenders were produced by BTI (Bubble Technology Industries)\footnote{%
Let us notice that they are no longer in production and have been
replaced by similar devices. However on the BTI there still is a web
page dedicated to them~\cite{def}.} and consisted of minute droplets
of a superheated liquid dispersed throughout an elastic polymer gel.
When neutrons strike these droplets, they form small gas bubbles
that remain fixed in the polymer. The number of bubbles is directly
related to the amount and the energy of neutrons, so the obtained
bubble pattern provides an immediate visual record of the neutron
dose (see Fig.\ref{neutr-ion}). Each Defender was provided with its
own calibration number ($bubble$ $number/mRem$), so it was possible
to convert the number of bubbles into dose equivalent.

We shall present two sets of experiments during which two different
kinds of bubble detectors were used: Defender and Defender XL. Their
specifications are slightly different. The Defender are cylinders
19.4 $cm$ long (their active part is 10 $cm$ long) and with a
diameter of 2.1 $cm$. Their minimum detection level is a tenth of an
ounce of Plutonium in seconds at 1 meter.
It was determined to be about 100 counts/$\mu Sv$ to $^{252}$Cf at 20 $%
{{}^\circ}%
C$. The Defender XL \ are cylinders 47 $cm$ long (their active part is 30 $%
cm $ long) with a diameter of 5.7 $cm$. They have a sensitivity
higher (by one order of magnitude) than that of the Defenders, their
minimum detection level being a hundredth of an ounce of Plutonium
in seconds at 1 meter. Their response was determined to be about
1000 counts/$\mu Sv$ to $^{252}$Cf
at 20 $%
{{}^\circ}%
$C. Both types of Defenders are sensitive to neutrons in the energy
range between 10 $keV$ and 15 $MeV$, and their response is dose
rate-independent and spatially isotropic. Moreover, both Defender
and Defender XL are completely unaffected by gamma radiation, as
stated by the manufacturer and
it was experimentally checked by irradiating them with a known source of $%
^{60}$Co for several minutes without producing the tiniest bubble.
Due to their very operation way, both kinds of Defender detectors
are sensitive to ionizing radiation too. However, the morphology and
the distribution of the bubbles are quite different, as shown by
Fig.\ref{neutr-ion} (supplied by the manufacturer, BTI). In the
picture on the left one sees the effect produced by neutrons in two
Defenders. The bubbles are big and spread out the whole volume.
Conversely, the two detectors on the right show a different type and
a different distribution of bubbles (generated by heavy ions),
gathered in a cluster and much smaller than those produced by
neutrons.

The CR-39 (PADC) plastic track detector is a C$_{12}$H$_{18}$O$_{7}$
polymer with density 1.3 $g/cm^{3}$ which is used for registration
of heavy charged particles and is a very convenient mean of
detection. Charged particles are registered directly, and neutrons
are detected through a secondary recoil particles or nuclear
reactions. The CR39 energy range sensitivity is very wide, from tens
of $keV$ to hundreds of $MeV$. Particle tracks on the detector
become visible after chemical etching and are investigated using a
microscope. In order to detect neutrons by the CR39, we used the
nuclear reaction $^{10}B$($n$,$\alpha $)$^{7}Li$ and hence spread a
2 $mm$ layer of natural Boron (80.1\% $^{11}B$, 19.9\% $^{10}B$) on
the CR39 detecting surface which had to convert neutrons into alpha
particles, following a well known
technology~\cite{tommasino,khan,izer}.

Measurements of ionizing ($\alpha $, $\beta $ and $\gamma $)
radiation background were carried out by using three types of
detectors: Geiger
counter Gamma Scout~\cite{scout} with a mica window transparent to $\alpha $%
, $\beta $ and $\gamma $ radiation, and provided with two aluminium
filters
1 $mm$ and 3 $mm$ thick, to screen $\alpha $ radiation and $\alpha $ and $%
\beta $, respectively; polycarbonate plate detectors PDAC CR39
sensitive to ionizing radiation in the energy range 40 $keV$ -4
$MeV$ and Tallium (Tl)
activated, Sodium Iodine (NaI), $\gamma $-ray spectrometer GAMMA 8000~\cite%
{amptek}.

\subsection{Experimental Results}

\subsubsection{First Investigation:Thermodynamical Detectors}

Two separate investigations have been carried out. In the first one,
we subjected to cavitation not only bidistilled deionized water, but
also
solutions of four different salts in H$_{2}$O (with concentration of 1 $ppm$%
), namely:

\begin{itemize}
\item 250 $ml$ of H$_{2}$O;

\item 250 $ml$ of H$_{2}$O solution of Iron Chloride FeCl$_{3}$;

\item 250 $ml$ of H$_{2}$O solution of Aluminium Chloride AlCl$_{3}$;

\item 250 $ml$ of H$_{2}$O solution of Lithium Chloride LiCl;

\item 500 $ml$ of H$_{2}$O solution of Iron Nitrate Fe(NO$_{3}$)$_{3}$.
\end{itemize}

\begin{figure}
\begin{center} \
\includegraphics[width=0.8\textwidth]{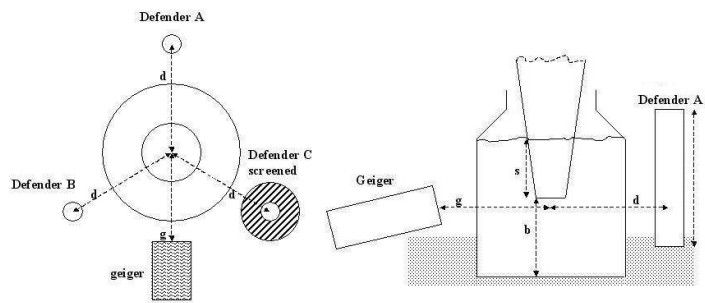} \caption{\textit{Layout and lateral
section of the experimental setup. d=7 cm, g=10 cm, s=4 cm, b=5 cm.
This setup indicates that between the cavitation area and the
neutron detectors and the Geiger counter there were 3.5 cm of water,
the thickness of the Borosilicate (about 2 mm) and few centimetres
of air.}}\label{layout}
\end{center}
\end{figure}

Each of the first four cavitations lasted 90 $min$, while the Iron
Nitrate solution was cavitated for 120 $min$. The schematic layout
of the experimental equipment is shown in Fig.\ref{layout}.

The cavitation chamber (vessel) was in the centre and the sonotrode
has to be imagined perpendicular to the plane of the figure, just
over the vessel and lined up with it. The immersion of the sonotrode
and the distance of its tip from the bottom of the vessel were about
4 $cm$ and 5 $cm$ respectively. They were suitably studied with
reference to the cavitation chamber, in order to maximize the
concentration of energy utilized for cavitation (by taking advantage
also of the pressure waves reflected by the bottom of the cavitation
chamber toward the sonotrode tip, and reducing the energy dispersion
in the piezoconvective motions of the cavitated liquid).

For each cavitation experiment, we used three neutron detectors
Defender. They were placed vertically and parallel to the vessel or
the sonotrode axis, arranged as shown in Fig.\ref{layout}. One of
the Defenders was screened by immersing it in a cylinder of carbon
(moderator) 3 $cm$ thick. The Geiger counter was pointed towards the
area inside the vessel where cavitation took place. A second equal
arrangement of three Defenders and the vessel containing the same
uncavitated solution (blank), was placed in a different room and was
used to measure the neutron radiation background at the same time
when cavitation was taking place. The measurements of fast neutron
radiation carried out in the experiments with H$_{2}$O, Aluminium
Chloride and Lithium Chloride were compatible with the background level (20 $%
nSv$). On the contrary, in the second and the fifth experiment, with
Iron Chloride and Iron Nitrate respectively, the measured neutron
radiation was \emph{incompatible} with the neutron background level.

Let us remark that the appearance of bubbles in these detectors can
be brought about by other sources than neutrons. Since the droplets
are in a metastable state, they can be turned into bubbles as they
receive the correct amount of energy, which can be conveyed to them
by heat and mechanical compressions, just like ultrasounds. As to
heat, the Defender detectors are temperature-compensated and their
correct operation is
guaranteed in the range from 15 $%
{{}^\circ}%
C$ to 35 $%
{{}^\circ}%
C$. Besides, the temperature of the laboratory (a small room) was
kept
constant at about $\left( 20\pm 1\right) $ $%
{{}^\circ}%
C$ by a heat pump that could work in reverse mode as well. Of course
we monitored by an infrared thermometer the temperature of the
Defenders all over their body and with particular care on the area
nearer to the vessel that became warm during cavitation. The
temperature of this specific part
never exceeded 26 $%
{{}^\circ}%
C$ , which is well within the working temperature guaranteed by the
manufacturer. By comparing the number of bubbles popped up during
each experiment, one can unmistakably state that they cannot be
brought about by heat, since the temperature increases of the
solutions treated by ultrasounds in all the experiments were always
compatible with each other
within $\pm $5 $%
{{}^\circ}%
C$, while the number of bubbles ranged from less than 10 up to 70.
As to the second possible source of bubbles, i.e. ultrasounds, the
ultrasound power and the experimental setup were the same for all
the five experiments, but
only in two out of five we got a neutron signal higher than the background%
\footnote{%
The neutron background measurements were carried out at the same
time of the cavitation, but in a different room, by means of equal
detectors placed around a similar vessel containing the same
solution. The results obtained were compatible with the background.
The same compatibility was found with detectors immersed in carbon
both in presence and in absence of cavitation. This last result is a
further confirmation of the neutronic origin of the bubble signals
in the Defenders.}. This evidence rules out ultrasounds as the
possible cause of bubbles in the Defenders.

\begin{figure}
\begin{center} \
\includegraphics[width=0.8\textwidth]{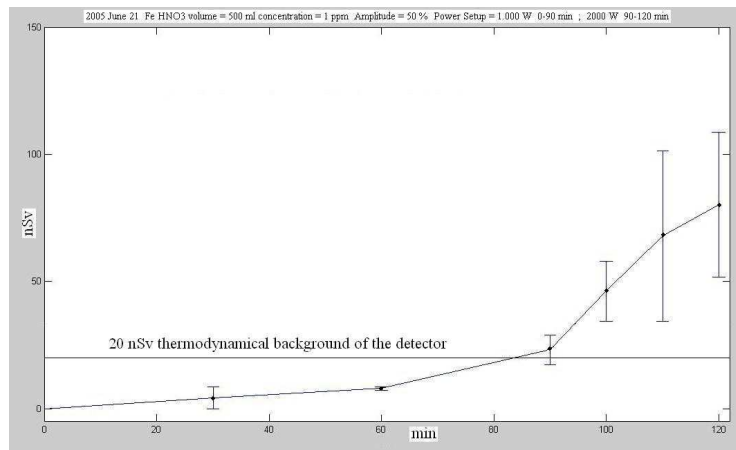} \caption{\textit{Neutron dose (nSv) vs.
cavitation time for Fe(NO}$_{3}$\textit{)}$_{3}$\textit{\ solution.
The horizontal line represents the background level.
}}\label{nitrate graph}
\end{center}
\end{figure}

A more quantitative behavior of the phenomenon is illustrated by Fig.\ref%
{chloride graph 1} and Fig.\ref{nitrate graph}, which report the
neutron doses in nano-Sievert as function of the cavitation time.
The horizontal black line represents the sum of the measured
thermodynamical instability of
the detectors\footnote{%
Indeed, when the detectors are activated one faces an initial
thermodynamical instability due to the almost sudden decrease of
pressure applied to the superheated droplets dispersed in the gel.
Some of them evaporate and form bubbles which have to be taken into
account as a background level of blindness of the detector beyond a
real, although very low, neutron background level.} and of the
measured neutron background level and is equal to 20 $nSv$. In both
graphs, the values correspond to the mean of the two equivalent
doses obtained by the two defenders without moderator
used during cavitation\footnote{%
The number of bubbles was visually determined by two of the
experimenters independently and the mean value of the two counts
(which were always absolutely compatible and almost always equal to
each other) was taken as the number of bubbles to calculate the
dose.}. The error bars were determined by taking the root mean
square of the differences of the two equivalent doses and the mean
value.

\begin{figure}
\begin{center} \
\includegraphics[width=0.8\textwidth]{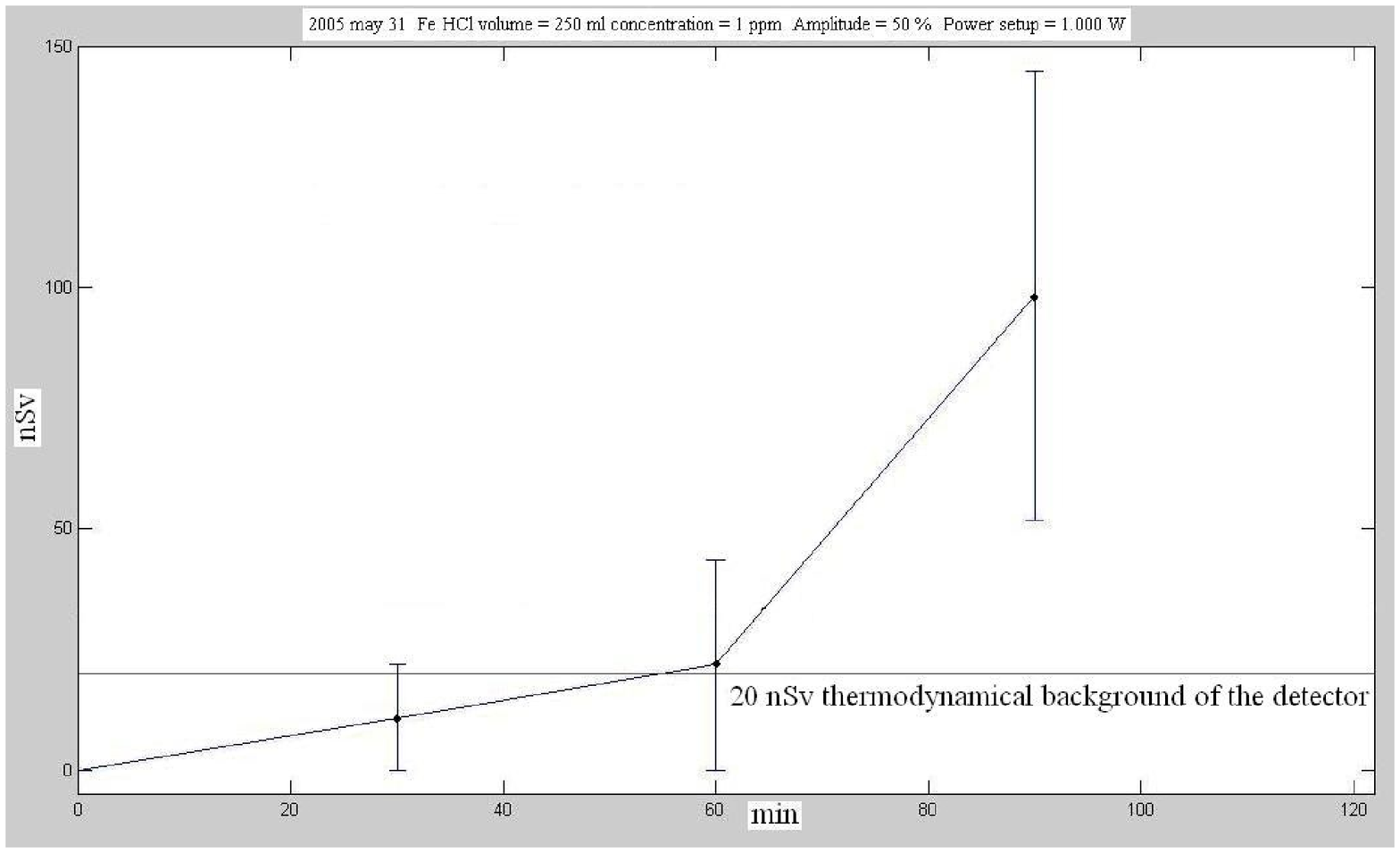} \caption{\textit{Neutron dose (}$\mathit{%
nSv}$\textit{) vs. the cavitation time for }$Fe(Cl)_{3}$
\textit{solution. The horizontal line represents the background
level.}}\label{chloride graph 1}
\end{center}
\end{figure}

In the last thirty minutes of cavitation of the iron salt solutions,
the measured dose (\symbol{126}100 $nSv$) was significantly higher
(even 5
times) than the background. Precisely, the final measured dose was (98.50 $%
\pm $ 4.5) $nSv$ for FeCl$_{3}$ (Fig.\ref{chloride graph 1})and
(76.00 $\pm $ 4.5) $nSv$ for Fe(NO$_{3}$)$_{3}$ (Fig.\ref{nitrate
graph}). Therefore, \emph{the neutron emission induced by cavitation
exhibits a marked threshold behaviour in time.}

The radiations $\alpha $, $\beta $ and $\gamma $, measured in all
the
cavitation runs, turned out to be compatible with the background radiation%
\footnote{%
This agrees with the results on the absence of radiation emission in
the first cavitation experiments~\cite{carmig1,carmig2}. See
Sect.2.} \newline

Let us comment the above results. The negative outcomes of the three
experiments with water, aluminium chloride and lithium chloride
allow one to conclude that the neutrons emitted during the
cavitation of iron chloride cannot be related to the presence of
$H_{2}O$ and $Cl$ in this experiment. As to the positive outcomes of
the fifth experiment with iron nitrate, although it cannot be
excluded that the neutron emission be related to the presence of
nitrogen, we can certainly state that this emission took place when
iron was part of the cavitated solution.

Moreover, the absence of ionizing radiation $\alpha $, $\beta $ and
$\gamma $ above the background level in all the experiments --- even
in those two in which we got the evidence of neutron emission ---
means that neutrons were produced without the usual consequent
emission of gamma radiation.

One can therefore state that \emph{only the presence of iron in the
cavitated solution gave rise to neutron emission (and therefore to
nuclear
processes induced by cavitation)}, but without the accompanying emission of $%
\gamma $ radiation. Of course, this could mean either that gamma
radiation was not emitted at all as it usually is when neutrons are
emitted, or that the sensitivity of our detectors was not sufficient
to reveal their slight presence. Besides, we have to point out that
even if neutron emission took
place without any consequent gamma radiation\footnote{%
A possible explanation of this fact, based on a space-time
deformation of the interaction region between two
nuclei\cite{carmig7}, will be discussed in Sect.7.} from nuclei
de-excitation, one would expect gamma rays to be emitted from
neutron capture by the proton of Hydrogen anyway.This is an issue
which will certainly be addressed in future experiments.

Moreover, the increase of the derivative that appears quite
evidently in the last 30 minutes of cavitation for FeCl$_{3}$
(Fig.\ref{chloride graph 1}) and Fe(NO$_{3}$)$_{3}$
(Fig.\ref{nitrate graph}). may be read as a first corroborating
evidence for the phenomenological model proposed by two of us (F.C.
and R.M.)~\cite{carmig7}, which assumes the existence of a threshold
in power and energy (and hence time) for piezonuclear reactions to
happen. In this sense, provided the ultrasonic power transmitted
into the solution is higher than the required
threshold~\cite{carmig7}, the emission of neutrons produced by these
reactions begins only after that a certain amount of energy was
conveyed into the solution or, equivalently, after a certain time
interval. This will be discussed in detail in Sect.7.

Further, if the bubble collapse is the main microscopical mechanism
to induce piezonuclear reactions (and hence neutron radiation)
\cite{carmig7}, then it is expected on physical grounds that the
emission of neutrons does not take place as from a stable source
but, conversely, it happens in bursts (as it occurs for
sonoluminescence \cite{RMP}). However, this consideration can be
considered at this stage only as a heuristic hypothesis. We shall
come back to this point in the following.

\subsubsection{Second Investigation:Enhanced Thermodynamical Detectors}

\begin{figure}
\begin{center} \
\includegraphics[width=0.8\textwidth]{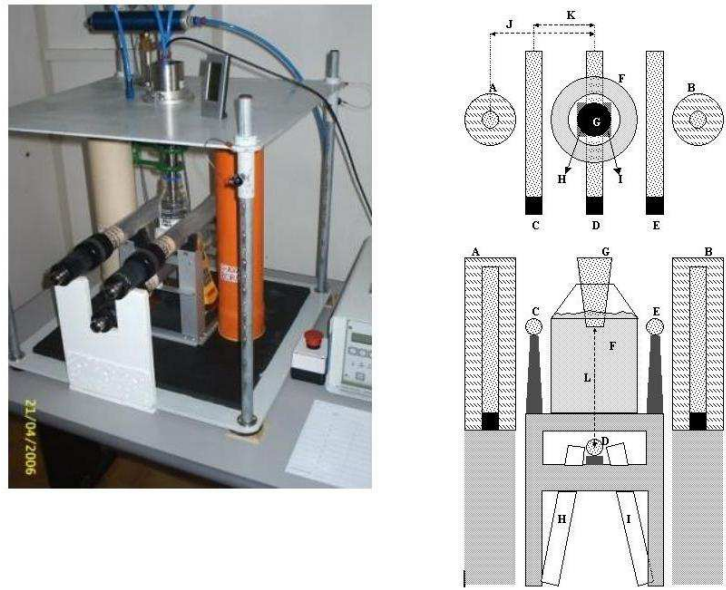} \caption{\textit{Experimental apparatus
used in the second investigation. The cavitation Chamber (F) is
visible in the middle of both pictures and the sonotrode, the
vertical tapered metal stick (G), is aligned with and inserted in
it. The green pipe surrounding the sonotrode conveyed the cooling
air onto the sonotrode surface. The three horizontal greyish
cylinders (C,D,E) with a black cylindrical endcap are the neutron
detectors. The two orange (right) (B) and creamy (left) (A) vertical
cylinders contained the two screened Defenders, one by boron
(orange) and the other by carbon (creamy). }}\label{2 layout}
\end{center}
\end{figure}

Since the first investigation highlighted the basic role of Iron in
producing piezonuclear reactions, the second one was devoted to a
systematic study of such an evidence, by using solutions with only
Iron Nitrate, since it gave rise, in the previous investigation, to
the maximum flux of emitted neutrons. Then, six cavitation runs
(each lasting 90 $min$) were carried out
on the same quantity (250 $ml$) of pure water and of a solution of Fe(NO$%
_{3} $)$_{3}$ with different concentration, subjected to ultrasounds
of different power. Namely, the cavitated solutions could have three
possible concentrations, 0 $ppm$ (H$_{2}$O), 1 $ppm$ and 10 $ppm$.
Moreover, the oscillation amplitude and hence the transmitted
ultrasonic power took two different values, 50\% and 70\%,
corresponding to about 100 $W$ and 130 $W$, respectively. The energy
delivered to the solution within the whole cavitation time was 0.54
$MJ$ and 0.70 $MJ$ in the two cases. In order to measure neutron
radiation we employed five neutron detectors of the Defender XL
type. Background neutron measurements were accomplished at the
beginning of the whole set of cavitations. During each cavitation we
carried out
ionizing radiation measurements by two Geiger counters Gamma Scout~\cite%
{scout} , one with no aluminum filter and the other with a 3 $mm$
filter, used simultaneously. A picture and the layout of the
experimental apparatus used in the six cavitation runs are shown in
Fig.\ref{2 layout}.

The two vertical cylinders (A and B) contained one Defender XL each,
of the same type of the three horizontal ones. The detectors were
surrounded, and hence screened, by 3 $cm$ of Boron powder (B)
(thermal neutron absorber) and by 3 $cm$ of Carbon powder (A)
(neutron moderator), respectively. In all of the six experiments of
this second investigation, the three horizontal, unscreened Defender
XL's measured a neutron emission significantly higher than the
background level. The two vertical, screened Defender XL's (both by
boron and carbon) always detected a reduced neutron dose, comparable
with the background one (thus again providing further evidence of
the neutron origin of the bubble signals).

\begin{figure}
\begin{center} \
\includegraphics[width=0.8\textwidth]{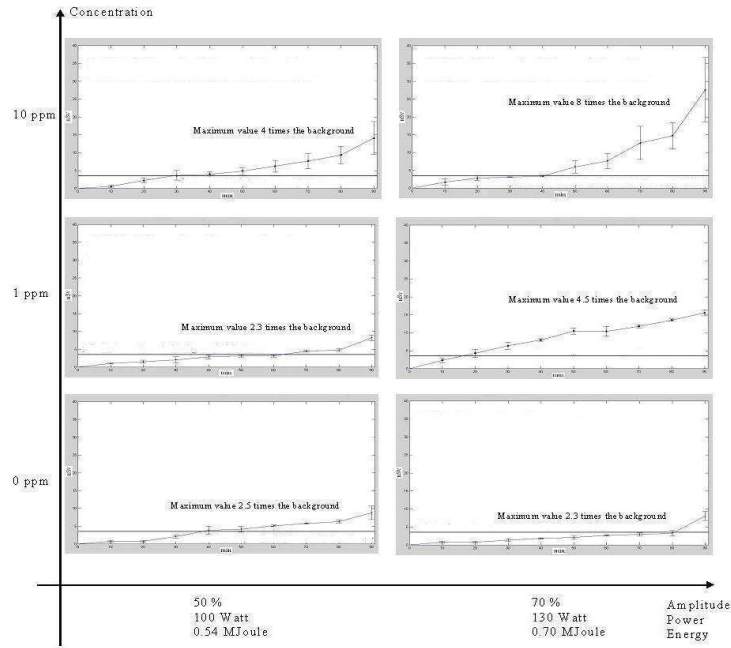} \caption{\textit{The six graphs (one for each cavitation of the
second series) showing the neutron dose (in }$\mathit{nSv}$\textit{)
as a function of time in minutes (time interval 10
}$\mathit{min}$\textit{). Each curve corresponds to one value of
concentration and one of the amplitude. The horizontal line in all
graphs corresponds to the thermodynamical background of 3.5
}$nSv$.\textit{\ The graphs are displaced in a Cartesian plane, with
concentration (in ppm) on the y-axis and amplitude (power) on the
x-axis.}}\label{graphs1}
\end{center}
\end{figure}

For all the six experiments, we plotted the measured doses of
neutrons (in nano-Sievert) as function of the cavitation time. The
number of bubbles was counted every 10 $min$. Each curve corresponds
to one concentration of the Fe(NO$_{3}$)$_{3}$ solution, from 0
$ppm$ to 10 $ppm$, and one oscillation
amplitude (and therefore ultrasonic power), 50\% (100 $W$) or 70\% (130 $W$%
). The six graphs are reported in Fig.\ref{graphs1}.

They are displaced in a Cartesian coordinate system with
concentration on the y-axis and amplitude (power) on the x-axis. As
in the first investigation, the horizontal black line represents the
sum of the measured thermodynamical instability of the detectors and
of the measured neutron background level. The six graphs of
Fig.\ref{graphs1} do not show the threshold behaviour in energy that
we found in the first investigation, namely the sharp and sudden
increase of the curve derivative in the last 30 minutes of
cavitation. Despite that, according to our heuristic hypothesis
(which will be experimentally supported in the following) about the
neutron emission taking place in bursts, it will be wrong to
interpret these curves as a sign of a stable neutron emission.
Conversely, still considering valid this hypothesis, one can ascribe
this different behaviour between the first and the second
investigations to the different immersions of the sonotrode in the
solution, which was about 4 $cm$ (about 5 $cm$ from the bottom of
the vessel) in the former and only 1 $cm$ (about 10 $cm$ from the
bottom of the vessel) in the latter. This means that both the height
of the neutron peaks (bursts) and, hence, the emitted dose can be
controlled somehow by this geometrical parameter. This consideration
allows one to ascribe this apparent lack of threshold behaviour to
the reduced height of the neutron peaks emitted during the
cavitations performed in the second investigation with respect to
those emitted in the first one. This reduced height spreads the
neutron dose over a longer period of time, thus preventing the
threshold behaviour from showing up. It will be the purpose of our
future investigations to establish the time of appearance of the
first neutron burst and verify whether it takes place beyond the
energy (or time) threshold.

Moreover, Fig.\ref{graphs1} further disproves the possible criticism
about a possible generation of the bubbles by ultrasounds rather
than by neutrons. Indeed, by looking at the compound graph and
reading it along its columns, i.e. keeping the amplitude (power)
fixed, it is seen that the curves are different, while the
ultrasonic power is always the same. Conversely, had ultrasounds
been the real cause of the bubbles, one should have had equal
effects. A possible origin of the bubbles due to heath must be
discarded on the basis of the precautions we took as in the previous
investigation (stabilization of the laboratory temperature and check
of the temperature of the body of the two Defenders XL). As a
further proof against any possible influence of temperature or IR
irradiation on the number of bubbles in the Defenders, we checked
that at equal temperature of the solution in the vessel, and equal
ultrasonic power, the bubble distribution in the Defender XL did not
show any systematic concentrations (qualitatively and quantitatively
in term of number of bubbles) near the warmest part of the vessel
and in the surroundings where possible thermal gradients might have
had some effect on the stability of the detectors.

Let us remark that in the second investigation one got evidence for
neutron emission also in cavitating pure water, unlike the case of
the first one. This is obviously due to the higher sensitivity of
the detectors employed in
the second investigation \footnote{%
One has not to be surprised by the emission of neutrons from the
cavitation of bidistilled deionized water since a mass spectrometer
analysis of its content cleary shows that it contains every chemical
element.}. Such a result agrees with the indirect evidence for
neutron emission obtained in the first experiment of water
cavitation, in which the changes in concentration of the stable
elements occurred with a variation in neutron
number~\cite{carmig1,carmig2}.

A difference between the two series of cavitations is provided by
the different amount of neutron doses in the two cases. Actually (as
it can be
seen by comparing Fig.\ref{graphs1} with Figs.\ref{chloride graph 1}, \ref%
{nitrate graph}) the neutron doses yielded by cavitation in the
experiments of the first investigation are nearly one order of
magnitude higher than the neutron doses yielded by cavitation in the
second one (indeed, in the last cavitation run we got a maximum of
(28.0 $\pm $ 7) $nSv$). This fact is even more puzzling since one
would expect higher doses in those experiments carried out with a
greater amplitude and hence higher ultrasonic power. A possible
explanation can be found in the different immersion depth of the
sonotrode in the solution. Indeed, the immersion depth in the first
investigation, about 6 $cm$, was just chosen in order to get the
cavitation process more effective. However, this circumstance needs
further researches in new experiments, in which neutron emission be
studied as a function of the immersion depth of the sonotrode, by
investigating both how the sonotrode tip wears out (in connection
with its oscillation mode, first or second Bessel harmonic) and how
the piezoconvective motions of the cavitated liquid go on.

At the light of the above results, we can say that \emph{the
cavitating device behaves as an ultrasonic nuclear reactor. }As we
have already said, we performed measurements of the ionizing
radiation by means of the above mentioned (filtered and unfiltered)
Geiger counters. The measured radiation was always compatible with
the background level. As a further check of the absence of $\gamma $
radiation, we carried out, in absence of cavitation and
during cavitation of Iron Nitrate (70\% amplitude, concentration $>$10 $ppm$%
, duration 90 $min$), simultaneous measurements by means of the two
Geigers and through a tallium (Tl) activated, Sodium Iodine (NaI),
$\gamma $-ray spectrometer. We found again a perfect compatibility
between the background spectrum and that during cavitation both for
the two Geigers and for the NaI (Tl), $\gamma $-ray spectrometer (in
spite of the neutron signal with maximum of (9.1 $\pm $ 0.5) $nSv$
measured by the Defender XL's). Thus, the results of the second
investigation too provided evidence for the emission of anomalous
nuclear radiation, since neutrons were not accompanied by gamma
rays. The NaI(Tl) spectrometer allowed us to increase by several
orders of magnitude the accuracy and sensitivity of gamma ray
detection. Despite that, we need again to raise the question about
the lack of gamma rays from neutron capture by the proton of
Hydrogen which need to be addressed in future experiments.

The systematic analysis carried out by cavitating water solutions of
Iron Nitrate, for all of which evidence of neutron radiation was
gotten, shows that \emph{the phenomenon is perfectly reproducible}.
Moreover, we have been able, by changing the immersion depth of the
sonotrode tip, to reduce the
emitted neutron dose by one order of magnitude. This implies that \emph{%
neutron emission can be somehow controlled.}

\subsubsection{Further Check and Features of Neutron Emission:Polycarbonate
Detectors}

In the previous two investigations, the evidence for neutron
emission was highlighted by means of the detectors Defender through
the analysis of the bubble signals. As a further check, we carried
out a further experiment utilizing not only the Defender XL's but
also boron-screened CR39 detectors.

\begin{figure}
\begin{center} \
\includegraphics[width=0.8\textwidth]{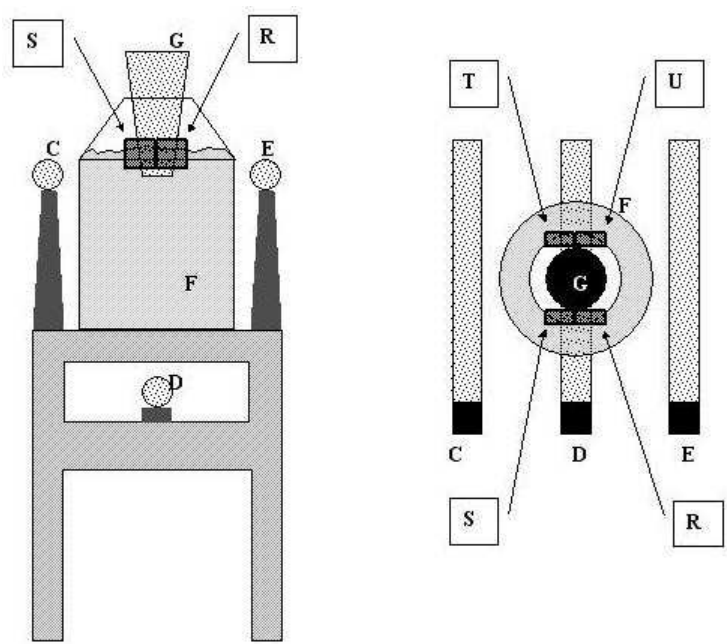} \caption{\textit{Layout of the
experimental set-up of the second investigation showing the position
of the Boron screened CR39 plates with respect to the rest of the
equipment.}}\label{CR39}
\end{center}
\end{figure}

By the same experimental apparatus used in the second investigation (see Fig.%
\ref{2 layout}), we subjected to cavitation 250 $ml$ of a water
solution of Iron Chloride (FeCl$_{3}$) with concentration 10 $ppm$.
The cavitation lasted 90 $min$ at the ultrasound frequency of 20
$kHz$, with oscillation
amplitude of 70\% of the maximum amplitude, corresponding to a power of 130 $%
W$ (namely to a total energy of 0.70 $MJ$). The choice to use again
a solution of FeCl$_{3}$ was due to the fact that, all the other
conditions being equal, we noted that with Iron Chloride there is a
higher release of macroscopic energy than with Iron Nitrate (the
liquid evaporation is from 2 to 5 times that observed with the
latter solution). Due to the equality of thermodynamical conditions,
this cannot be explained in terms of ultrasounds only. The two
unscreened lateral Defender XL's (C and E) measured a maximum dose
of neutrons of 14.5 $nSv$, 4 times higher than the detector
thermodynamic noise of 3.5 $nSv$. Moreover, we placed, externally to
the cavitation chamber, two pairs of 1 $cm$ by 1 $cm$ plate CR39
detectors (R,S and T,U) as shown in Fig.\ref{CR39}.

\begin{figure}
\begin{center} \
\includegraphics[width=0.8\textwidth]{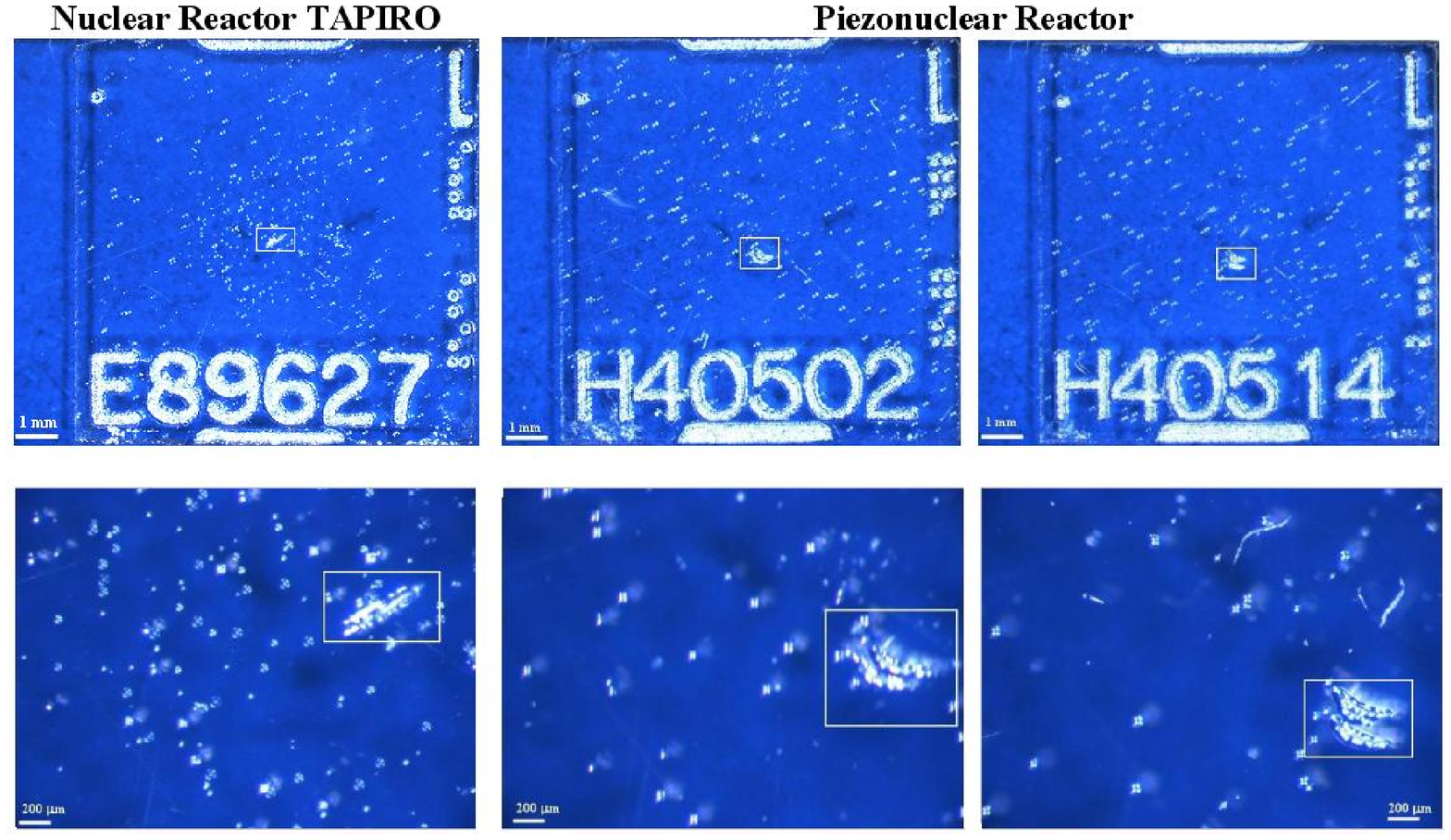} \caption{\textit{The first column shows
the Boron-CR39 detector irradiated by the nuclear reactor TAPIRO at
Casaccia ENEA with a neutron dose of 2.1 }$\protect\mu
\mathit{Sv}$\textit{\ (upper row: magnification X 10 - lower row:
magnification X 50), the second and third columns show the two
Boron-CR39 detectors that were next to the cavitation chamber during
the application of ultrasounds (upper row: magnification X 10 -
lower row: magnification X 50).}}\label{crtraces}
\end{center}
\end{figure}

Each plate was at a distance of about 4 $cm$ from the vertical axis
of the cavitation chamber, at the same level of the sonotrode tip.
In between the
CR39 plates and the axis of the vessel there were 3.5 $cm$ of solution, 2 $%
mm $ of borosilicate glass and about either 3 $mm$ of air or 3 $mm$
of Boron. The two couples were diametrically opposite to each other.
In each pair, a CR39 was in air (S and T), whereas the other
detector was immersed in boron (R and U) (whose interaction with
neutrons gives rise to alpha radiation to which CR39 are sensitive).
The results obtained are displayed in the second and third row of
Fig.\ref{crtraces}. By the boron CR39 we were able to detect
neutrons with energies below 10 $keV$ too and, above all, thermal
neutrons.

In order to have an idea of what the traces should look like on
these detectors after etching, four more detectors were irradiated
by neutrons using as source the fast neutron nuclear reactor TAPIRO
at Casaccia ENEA Rome. The neutron equivalent dose conveyed onto the
detectors was 2.1 $\mu
Sv $ through a diagnostic neutron channel\footnote{%
Not knowing what kind of neutron spectrum to expect from the
cavitated solution, as already stated, we decided to produce our
comparison model of traces by a source whose spectrum were the
widest possible, i.e. a nuclear reactor. According
to~\cite{tommasino} these kind of detectors can detect fast,
epithermal and thermal neutrons, of course with different
sensitivities . Hence the integral effect on the detectors, due to
almost the whole neutron spectrum, would be traces whose quantity
and shape have to be compared with those obtained from the
piezonuclear reactor. Let us repeat that the main target of these
investigations was to reveal the presence of neutrons in a sort of a
'yes or no' detecting procedure.}. A boron CR39 was used to measure
the background level around the reactor, other two, one in air and
the other immersed in boron, were placed at about 3 $m$ from the
reactor core and radiated for 5 $min$. The results are shown in the
first column of Fig.\ref{crtraces}. The Boron-CR39 with code E89627
was placed with its detecting surface orthogonal to the neutron
channel of the nuclear reactor TAPIRO with its centre on the channel
axis. It clearly shows the expected tracks at its centre (a 50
magnification photo is presented as well for a better vision). The
other three CR39 used to measure the background around the nuclear
reactor do not show any thick track and the number of tracks on them
is much lower than that on the E89627. The CR39 detectors with code
H40502 and H40514 were used during cavitation and show tracks
absolutely compatible with that produced by the neutrons of the
nuclear
reactor\footnote{%
The chemical etching of all the CR-39 foils was carried out in a
6.25N NaOH (Carlo Erba standard) solution at 90$^{\circ }$C for 4.5
hours according to the specifications given by the firm FGM Ambiente
which provided them.}.The comparison between the traces produced by
neutrons in the CR39 immersed in boron in the nuclear reactor case
(first column) and in the ultrasound one (second and third column)
shows that their pattern (although not their extension) is perfectly
similar. It is also possible to notice that the area of the thick
trace produced by the reactor neutrons is about half of the areas of
the thick traces produced by the neutrons generated by ultrasounds
during cavitation. The Boron-CR39 detectors can reveal neutrons of
any energy. While fast neutrons are not affected by Boron and leave
their own traces on the polycarbonate surface, slow neutrons and,
above all, thermal
neutrons, are converted into alpha particles by interacting with Boron-10 ($%
^{10}$B) (according to $^{10}$B(n,$\alpha $)$^{7}$Li) and through
this mechanism produce a much wider and deeper trace on the
polycarbonate surface than fast neutrons. By using this fact and
comparing the CR39 traces obtained in this experiment (compatible
with equivalent doses of 4-5 $\mu Sv$ in 90 minutes), with the
bubble signals collected by the Defender XL's in this same
experiment (14.5 $nSv$ in 90 minutes), and with those by the
Defenders of the first investigation (between 80 and 100 $nSv$ in 90
minutes), we are allowed to conclude that the bulk of the neutron
emission corresponds to neutrons having energy in the low epithermal
range and even lower. We believe that the outcomes shown by the
photos in Fig.\ref{crtraces} represent a fairly sound proof to
corroborate our heuristic hypothesis about the emission of neutrons
in bursts. Of course, the reactor channel acted as a filter which
selected those neutrons with velocity almost parallel to the channel
axis. In fact, the distribution of the tracks on the central part of
the plate E89627 is nearly circular around the thick track, implying
that the emission of neutrons from the reactor core was constant and
isotropic. This is the consequence of the cylindrical channel
through which it was irradiated, that produced a neutron flux with a
cylindrical symmetry. On the contrary, despite the cylindrical
symmetry of our experimental equipment (the vessel and the
sonotrode), it is quite clear that the neutron emission during
cavitation was neither constant nor isotropic. Were it isotropic,
one would have got a more uniform distribution of traces and more
thick traces on the CR39 plates and a more uniform distribution of
bubbles in the defenders. Conversely, the other two CR39 (H40502 and
H40514) do not show any particular distribution of tracks consistent
with the lack of any preferred direction of neutron emission from
the volume where cavitation was taking place. Despite that, on the
two CR39 detectors there are two thick tracks at the centre of the
chips which perpendicularly faced the centre of the cavitation
volume. This difference can be considered as a strong hint to state
that the neutron emission is not continuous but, conversely, takes
place in bursts at different instants of time, along diverse space
directions and with different height and energy spectrum (like is
the case for sonoluminescence). This is obviously due to the fact
that the microscopical mechanism underlying neutron emission is
bubble collapse, which is governed by quite a few variables (yet
uncontrolled), like bubble dimension, quantity and type of atoms on
the bubble surface. We got a definite proof of the emission of
neutrons in bursts in our third investigation.

\subsection{Third Investigation:Electronic Detectors}

Having achieved several positive evidences of neutron emission from
cavitation, we decided to refine our research by performing the same
measurements by a Boron Triflouride detector \cite{carcher2}.

\begin{figure}
\begin{center} \
\includegraphics[width=0.8\textwidth]{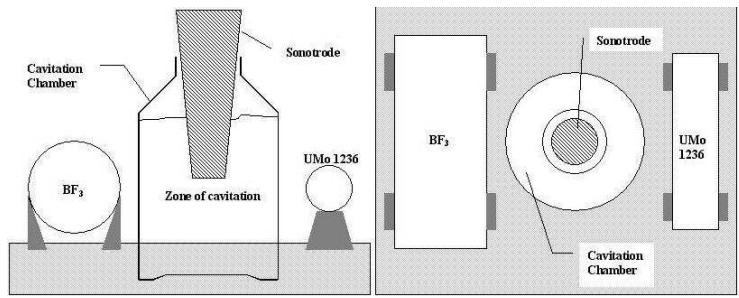} \caption{\textit{Vertical and horizontal layout of the experimental set-up.}}\label{lay-out}
\end{center}
\end{figure}

We cavitated 250 $ml$ of a solution of Iron Chloride FeCl$_{3}$ by
applying to it 130 stable $W$ of ultrasonic mechanical vibration at
20 $kHz$ for 90 minutes, employing the same sonotrode and the same
cavitation chamber of the precvious investigations. The immersion of
the sonotrode in the solution was
5 $cm$ \footnote{%
Further details about the geometry of the sonotrode and the
cavitation chamber and about their exercise are contained in three
patents owned by the Consiglio Nazionale delle Ricerche (CNR)
(National Council of Researches of Italy) now published
in~\cite{patent1,patent2,patent3}.}. During cavitation, we measured
neutron emission by a Wedholm Medical 2222A Boron Trifluoride
neutron monitor~\cite{wedholm} and ionizing radiation (in particular
gamma radiation) by an UMo LB 1236 monitor, whose energy range is
from 30 $keV$ up to 2 $MeV$ and which gives both the equivalent dose
and the equivalent dose rate in a wide range from 50 $nSv/h$ up to
10 $mSv/h$~\cite{berthold}. The UMo1236 was calibrated by a
Cobalt-60 standard source. The neutron monitor
was calibrated by an Americium-241 Beryllium standard source\footnote{%
This equivalent dose rate value is given by the manufacturer and
refers to a new source. The Americium-241 Beryllium neutron source,
that was used for the BF$_{3}$ calibration, was a four-year old
standard source which, considered that the Americium-241 half life
is of 432.2 years, can be considered as new.} contained within a
suitable lead and steel shielding box, in order to obtain outside of
it and in contact with it an equivalent
dose rate of neutrons of $\left( 1.5\pm 0.2\right) $ $\mu Sv/h$ . The BF$%
_{3} $ neutron monitor was placed in contact with the shielding box
containing the AmBe source, and its position was such to present to
the neutron flux a geometrical effective surface of 142 $cm^{2}$
perpendicular to it. We registered the number of counts from the
BF$_{3}$ within 100 seconds. Many of these counting runs were
carried out (sample size $\geq $
30) and we found out that the mean number of counts within 100 $s$ was 100 $%
\pm $ 15 (mean $\pm $ standard deviation) i.e. about 1 \ per second (1 $Hz$%
). The electronic noise value 0.03 $\cdot $ 10$^{-3}$
$counts/(s\cdot ~cm^{2})$ and the background flux 0.1 $\cdot $
10$^{-3}$ $counts/(s\cdot
~cm^{2})$ were summed, and the result 0.13 $\cdot $ 10$^{-3}$ $%
counts/(s\cdot ~cm^{2})$) was adopted as the error to be attributed
to each measured value. The layout of the experimental set-up is
schematically presented in Fig.\ref{lay-out}

\begin{figure}
\begin{center} \
\includegraphics[width=0.8\textwidth]{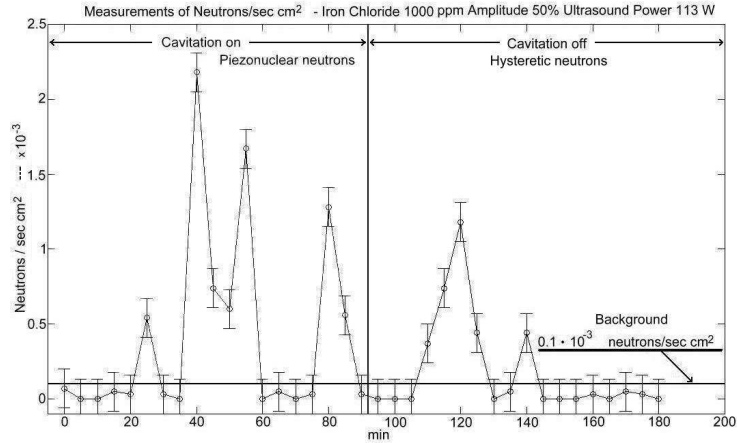} \caption{\textit{The
graph shows the neutron pulses obtained during one of the cavitation
runs.
Time in minutes is on the x-axis and neutron flux (}$neutrons/s\cdot cm^{2}$%
\textit{) }$\cdot $\textit{\ 10}$^{-3}$\textit{\ is on the y-axis.
The error bars represent the sum of the pessimistic measured
electronic noise of the whole measuring equipment and the
pessimistic measured laboratory background flux i.e. 0.13 }$\cdot
$\textit{\ 10}$^{-3}$\textit{\ }$counts/(s\cdot
cm^{2}).$}\label{graphic}
\end{center}
\end{figure}

The BF$_{3}$ detector was placed next to the cavitation chamber and
its position with respect to the zone of cavitation where the
neutrons are expected to be emitted was the same it had with respect
to the AmBe source during the initial calibration tests. The gamma
meter UMo 1236 was next to the bottle, diametrically on the other
side with respect to the BF$_{3}$. In Fig.\ref{graphic} we present
an example of the neutron measurements carried out during
cavitation, i.e. the application of ultrasounds to the 1000 $ppm$
solution of Iron Chloride. The BF$_{3}$ measurement runs were
triggered and checked by a Defender XL (not shown in
Fig.\ref{lay-out}) measuring as well, and testing \ that the dose
measured by both was the same within the errors, The pulse
collecting procedure was identical to that used while measuring the
background flux, i.e. registering the number of pulses accumulated
within 5 minutes.

Let us analyse the results depicted in Fig.\ref{graphic}. The
graphic is divided into two parts by a vertical line (at 92 minutes
instead of 90 minutes for mere visual convenience) . The left side
from 0 to 90 minutes is the interval of time during which cavitation
was on. The right side from 90 to 180 minutes is the interval of
time during which cavitation was off, but the neutron measurement
went on. On both sides, some peaks stand well above the background
level, pointing out that \emph{the emission of neutrons is not
constant in time, but occurs in bursts of neutrons or better in
pulses, thus definitely confirming our heuristic hypothesis}.
Moreover, we performed a careful statistical analysis of the pulse
values collected in the different cavitation runs, in order to
verify that they were not normally distributed, by exploiting the
graphical method called normal probability plot. The values of the
flux of the neutron pulses showed in no way to have a Gaussian
distribution. This shows that the neutron pulses are not equal
pulses emitted at different times in different directions, but
\emph{they are different} indeed.

In the left side of Fig.\ref{graphic}, the first neutron pulse
occurs after 40 minutes from the beginning of the cavitation and
this circumstance was the same for all the cavitation runs. More
precisely, it turned out that in all cavitation runs the first
neutron pulse appeared 40 - 50 minutes after switching on the
ultrasounds. \emph{This is a further confirmation of the threshold
behaviour in time (and therefore in energy) of the neutron emission.
}As to the right side of Fig.\ref{graphic}, although the cavitation
was turned off and hence one would expect that the neutron pulses
would stop along with it, there are two more peaks well above the
background level. These delayed pulses, emitted in all cavitation
runs about 20 minutes after the cavitation had been stopped, are a
hysteretic behaviour. A possible candidate explanation to this fact
is that some of the piezonuclear acoustic neutrons (those emitted
during cavitation) had been absorbed most likely by the Carbon
contained both in the steel sonotrode and in the materials of the
supporting platforms, and released after a latency (of about 20
minutes in our case), as it normally occurs in the graphite of
nuclear reactors.

Before drawing some conclusions, it is worth stressing some points
that, in our opinion, are important in order to create as a clear
and thorough picture of this new phenomenon as possible. As is well
known, there exists a
phenomenon in solids called fracto-emission or fracto-fusion~\cite%
{kaushik,storms} during which neutron bursts can be detected.
However, in most cases the cracked solids are loaded with deuterium,
which is not our case. Moreover, that our results have nothing to do
with fracto-emission has been proved by the fairly large number of
cavitation runs that we carried out under different conditions. Let
us recall (see Subsect.3.2) that we cavitated Aluminium Chloride and
Lithium Chloride solutions applying to them the same ultrasonic
power (i.e. the same oscillation amplitude of the sonotrode for the
same amount of time) without registering any neutron peak like those
in Fig.\ref{graphic}, so the level of the flux was always compatible
with the background. Besides, we performed different cavitation runs
of the same kind of solution both without changing the sonotrode and
by a new sonotrode for each run, in order to test possible effects
due to the sonotrode aging and damaging. No correlation of aging and
damaging with neutron emission was found, allowing us to exclude any
possible implication of fracto-emission.

The second experimental fact that we want to highlight is related to
the association between cavitation and microscopical thermonuclear
fusion occurring at the bubble
collapse~\cite{tal1}-\cite{tal4},\cite{storms}. As already stated,
the solutions that we used and from which we obtained neutrons did
not contain any Deuterium, but only deionized bidistilled water with
some Iron Chloride or Iron Nitrate. Besides, it seems impossible to
think of H$_{2}$-H$_{2}$ fusion, since the two solutions of
Aluminium Chloride and Lithium Chloride did not produce any sign of
neutrons above the background level, although the first two had the
same concentration of that with Iron and all of them were treated
with the same ultrasonic power for the same amount of time. The rest
of the stands and racks that were used to build the experimental
set-up were made of Iron, Aluminium, Polychloroprene (Neoprene),
Polymethyl Methacrylate (Plexiglas) that being sold for the most
different applications are not made of unstable radioactive or
fissile elements or even less contain large quantities of light
elements like Deuterium. Of course, it goes without saying that all
the neutron and gamma laboratory background measurements were
intended both to fix a level of blindness for the values that we
were going to collect and also to double check the remote chance of
any sort of tiny activity of the above materials. As stated above,
ionizing radiation detection was carried out in parallel with
neutron detection. In particular, gamma rays were monitored by the
UMo 1236 detector. The neutron and gamma monitoring proceeded
constantly in parallel for 180 minutes. The gamma response was both
in equivalent dose rate and in equivalent dose. The variations of
the gamma equivalent dose rate over 180 minutes were compared with
the variations of the $cps/cm^{2}$ from the neutron detector.
Particular attention was paid in order to make out possible gamma
peaks around those times when a neutron pulse occurred. Neither
coincidence nor correlation was found between neutron pulses and
gamma equivalent dose rate and dose, which turned out always
compatible with the gamma background, whose variations (of the order
of $\left( 0.14\pm 0.05\right) $ $\mu Sv/h$ (mean $\pm $ standard
deviation) for equivalent dose rate and $\left( 0.22\pm 0.07\right)
$ $\mu Sv$ (mean $\pm $ standard deviation) for equivalent dose) had
been extensively studied all over the laboratory.

Another point we want to stress is that in the experiments on liquid
solutions (see Sect.3), aluminum atoms appeared at the end in a
final quantity as large as about seven times the small initial
quantity. More precisely, they increased from 3.99 $ppb$ to 27.70
$ppb$, without evident reasons, if we exclude possible
transmutations. Therefore, our conjecture is
that the following piezonuclear fission reaction should have occurred:%
\begin{equation}
Fe_{26}^{30}\rightarrow Al_{13}^{14}+2\text{ }neutrons.
\end{equation}%
We shall come back to this point in Sect.5.

\section{Piezonuclear Reactions in Solutions of Unstable Elements: The
Thorium Experiment}

In all our experiments, described in the previous Section, the
cavitated elements were stable. However, at the light of the results
of the Russian teams, who observed the alteration of secular
equilibrium of Thorium 234 due to the explosion of titanium foils in
a solution of uranyl sulfate in distilled water
\cite{uru1}-\cite{uru3}, we decided to check the possible effects of
cavitation on the decay of an unstable nuclide. We chose as
radioactive nuclide thorium $Th^{228}$ and subjected to cavitation a
thorium solution in water \cite{carmig5}.

The employed equipment was the same used in the previous cavitation
experiments (see Subsect.3.1).

We prepared 12 identical solutions of $Th^{228}$ in pure deionized
bidistilled water (18 $M\Omega $), with volume of 250 $ml$ and
concentration ranging from 0.01 to 0.03 $ppb$ (part per billion).
$Th^{228}$ is an unstable element whose half life is $t_{1/2}$= 1.9
$years$ = 9.99$\times 10^{5}$ $min$. It decays by emitting 5 $\alpha
$ and 2 $\beta ^{-}$. The minimum energy of the alpha particles
emitted is 5.3 $MeV$, which is nearly equal to the energy of the
$\alpha $'s emitted by Radon 222. This alikeness allowed us to use
the detector CR39, a polycarbonate chip commonly used to detect
$\alpha $ particles emitted by Radon 222.

Eight solutions out of the twelve at our disposal were divided into
two groups of four, and each of them was cavitated for $t_{c}$ = 90
$min$ at a frequency of 20 $kHz$ and a power of 100 $W$. The surface
of the liquid was free. The remaining four were not cavitated, and
regarded as reference solutions.

We measured the ionizing radiation in the empty Duran vessel both
before and after cavitation. The radiation measurements were carried
out by means of two Geiger counters with mica windows (one of which
equipped with an
aluminium filter 3 $mm$ thick), and of a tallium activated, sodium iodine $%
\gamma $-spectrometer. The results turned out always compatible with
the background level. For each cavitation run, a CR39 detector was
placed inside
the vessel, on its bottom, and exposed for the whole cavitation time of 90 $%
min$.

The 12 detectors CR39 corresponding to the 12 solutions were
examined. The traces on them were clearly recognized as produced by
the $\alpha $ radiation from $Th^{228}$ decay, on a double basis.
First, such a trace has a characteristic, unmistakable "star-shaped"
look, completely different from
those impressed on CR39 by environmental radioactivity (\textit{e.g. }$%
Rn^{222}$) and by cosmic rays (as well known from the use of CR39
counters in environmental dosimetry).\ Furthermore, as a further
check, we inserted the impressed CR39 plates in the automatic
counter system "Radosys", which stated the incompatibility of our
traces with those of its database (just based on $Rn^{222}$ and
cosmic rays).

\begin{figure}
\begin{center} \
\includegraphics[width=0.8\textwidth]{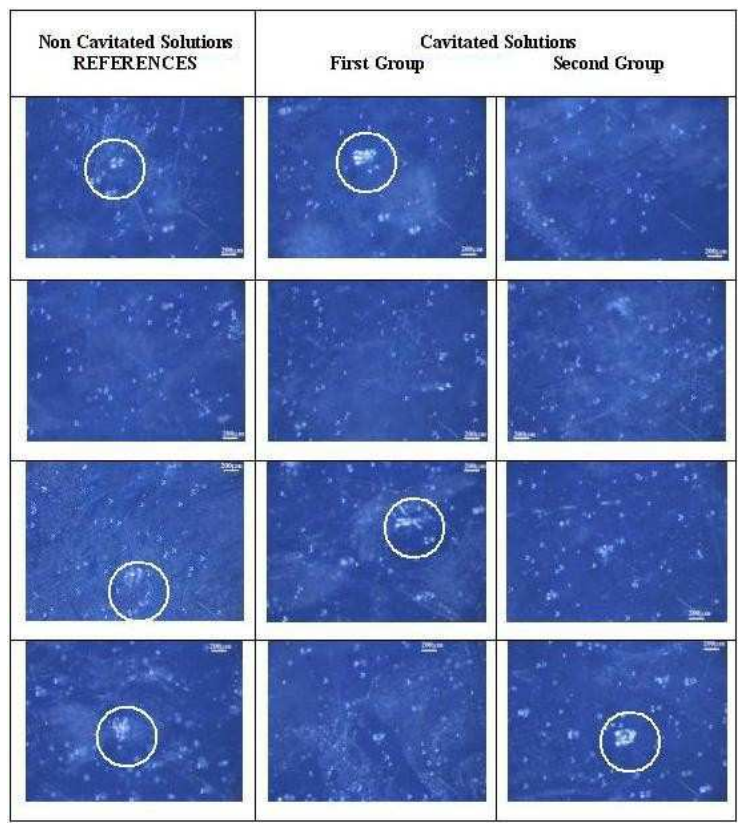} \caption{\textit{Traces left by }$%
\protect\alpha $\textit{-particles emitted from thorium decay on
detectors CR39 (circles).}}\label{traces}
\end{center}
\end{figure}

The results obtained are depicted in Fig. \ref{traces}. Precisely,
the first column shows the four detectors CR39 used with the four
non-cavitated solutions taken as reference, whereas in the second
and third columns one sees the eight detectors used with the eight
cavitated solutions. The circles in the figure highlight the traces
left by the particles $\alpha $ produced by thorium decay, which
were counted. On the four CR39 used with the four reference
solutions one counted 3 traces of alpha particles in all. The same
number of traces was counted on the eight CR39 used with the eight
cavitated solutions. Of course, in absence of any anomalous
behavior, one would have expected the same number of events for
either uncavitated and cavitated solutions.

On the contrary, the ratio of the number of traces and the number of
solutions is therefore 3/4 for the reference solutions and 3/8 for
the cavitated ones. Thus, there is an evidence of reduction of the
number of traces of alpha particles from thorium decay in the
cavitated solutions with respect to those in the non-cavitated ones.
In particular, it is evident that the above ratios show a reduction
by a factor 2 in the number of traces from the former with respect
to the latter.

In order to enforce the evidence obtained by the reduced statistics
of the
detector analysis, we analyzed by a mass spectrometer the content of $%
Th^{228}$ of all solutions, including those providing no evidence of
alpha particles from thorium decay. This was done by taking 40 $ml$
for each solution, on which we carried out 4 mass-spectrometric
analyses with a
drawing of 20 $\mu l$ and a scanning time of 150 $s$. The content of $%
Th^{228}$ (both in $ppb$ and in counts per second, $cps$) found in
the three cavitated solutions (whose CR39 showed the traces of the
alpha particles emitted by thorium) is half of that in the three
reference solutions corresponding to an $\alpha $-emission. This
situation is reported in Tables 1 and 2. Let us notice that the
samples which produced no signal in the CR39 counters did however
show the same halving of thorium content.

\begin{center}
\begin{tabular}{|c|c|c|}
\hline\hline
\multicolumn{3}{||c|}{\textbf{Table 1 }- \textit{Content of }$Th^{228}$%
\textit{\ in non-cavitated (reference) solutions}} \\ \hline\hline
\textbf{Mass-spectrometer analysis} & $cps$ & $ppb$ \\ \hline
\textbf{Sample 1} & 287$\pm 1$ & 0.020$\pm 0.01$ \\ \hline
\textbf{Sample 3} & 167$\pm 1$ & 0.012$\pm 0.01$ \\ \hline
\textbf{Sample 4} & 363$\pm 1$ & 0.026$\pm 0.01$ \\ \hline
\textbf{Mean values} & 272.33 & 0.019$\pm 0.01$ \\ \hline
\end{tabular}

\bigskip

\begin{tabular}{|c|c|c|}
\hline\hline
\multicolumn{3}{||c|}{\textbf{Table 2 }- \textit{Content of }$Th^{228}$%
\textit{\ in cavitated solutions}} \\ \hline\hline
\textbf{Mass-spectrometer analysis} & $cps$ & $ppb$ \\ \hline
\textbf{Sample 1 (first group)} & 231$\pm 1$ & 0.016$\pm 0.01$ \\
\hline \textbf{Sample 3 (first group)} & 57$\pm 1$ & 0.004$\pm 0.01$
\\ \hline \textbf{Sample 4 (second group)} & 79$\pm 1$ & 0.006$\pm
0.01$ \\ \hline \textbf{Mean values} & 122.33 & 0.009$\pm 0.01$ \\
\hline \textbf{Ratio of mean values non-cavitated/cavitated} & 2.2 &
2.1 \\ \hline
\end{tabular}
\end{center}

These two converging evidences allow one to conclude that\emph{\ the
process of cavitation reduced the content of }$Th^{228}$\emph{\ in
the solutions.}

The dry residues of both the cavitated and uncavitated samples have
been examined by X-ray microanalysis by means of an electronic
microscope (because it was impossible to insert them in a mass
spectrometer). However, this did not allow us to determine the
thorium variation in a clear way.

The ratio between the half life of thorium, $t_{1/2}$ = 1.9 $years$ = 9.99$%
\times 10^{5}$ $minutes$, and the time interval of cavitation, $t_{c}$ = 90 $%
min$, is $t_{1/2}/t_{c}$ = 10$^{4}$. This means that
\emph{cavitation brought about the reduction of }$Th^{228}$\emph{\
at a rate 10}$^{4}$\emph{\ times faster than the natural radioactive
decay would do.}

However, it must be stressed that \emph{this does not mean at all
that the
radioactivity of Thorium has been increased by a factor 10}$^{4}$\emph{. }%
Had been this the case, one would have to observe a greater number
of traces on the CR39 placed in the $Th^{228}$\emph{\ }solutions
subjected to cavitation, that does not. The concomitant absence of
any variation of other ionizing radiations ($\gamma$) (and the
absence of neutrons) apparently rules out the fact that the effect
of cavitation on thorium was simply to accelerate its natural decay
process (although this is still an open question deserving further
investigations). The most probable interpretation is assuming that
thorium underwent other kinds of transformations of piezonuclear
origin, like nucleolysis or nucleosynthesis. Anyway, we can conclude
that our results do support the Russian findings about the
alteration of the secular equilibrium of thorium, and give further
evidence of piezonuclear effects.

Thanks to some comments on this experiment raised by two research groups\cite%
{svedesi,kowalski}, we had the chance to clarify some points and add
some more comments and remarks of our own\cite{reply sved,reply
kowa}. A remark was about the limited collection of experimental
evidences\cite{svedesi} and hence a limited statistical analysis.
Although this is true, it was not our aim to carry out a precise
inferential statistical analysis in order to ascertain with a very
low level of significance if the reduction of radioactivity observed
in the cavitated samples were due not to chance, but to cavitation
itself\cite{reply sved}. To our knowledge, this experiment is the
first one to treat by ultrasounds and cavitation radioactive
elements
and it goes along with the evidences of the Russian team\cite{uru1,uru2,uru3}%
. Actually, our paper must be inserted in the context of the
evidences from other analogous experiments, in which cavitation was
shown to induce \textquotedblleft anomalous\textquotedblright\
nuclear effects. A further remark was raised by pointing out the
failures of Marie Curie who, at the
beginning of the last century, attempted to accelerate the decay of Radium%
\cite{kowalski}. However, by carefully looking at the experimental
conditions where all of these anomalous effects (anomalous
transmutations, emission of neutrons from Iron, and all of the
evidences collected in 20
years of low energy nuclear reactions\cite%
{carmig1,carmig2,carmig3,carmig4,carcher1,carcher2,storms}) take
place and from some of the predictions of our
theory\cite{carmig6,carmig8}, one understands that these phenomena
need very precise local condition, in terms of energy density and
time of release of this energy, which are not easy at all to be
simultaneously achieved. From this perspective, the attempts to
speed up alpha emission carried out between the end of the XIX and
the beginning of the XX century were very unlikely to be
systematically successful because of the difficulty to make out
these effects among other more evident phenomena. Other remarks
hypothesized that the decreasing of Thorium was due to the
depositing of it on the lateral and bottom surfaces
of the cavitation chamber or even the lateral surface of the sonotrode\cite%
{kowalski}. This circumstance was rebutted by pointing out that the
geometry of the cavitation chamber induced convective flows that
prevented Thorium atoms or compounds from being adsorbed\cite{reply
kowa}. A further remark concerned the star shaped tracks on the CR39
and the impossibility to have this shape during
cavitation\cite{kowalski}. In\cite{reply kowa} it is possible to
find the details of the measurement procedure that explains why the
star shaped tracks are possible. Before moving to the next subject,
we would like to point out that a Canadian team\cite{canada wrong
exp} attempted to repeat our experiment but the experimental setup,
that they used, had nothing to do with ours and in\cite{reply canada
wrong exp} we explain why this experimental setup is completely
unsuitable to produce the anomalous effects that we have collected
so far like those about Thorium.

\section{Piezonuclear Reactions in Brittle Fracture of Solids}

In Sects.2-4\ we have seen that in our experiments\ \cite{carcher1}-\cite%
{carmig5} the pressure of ultrasonic waves in a liquid causes the
cavitation of the gasses dissolved therein, with the ensuing
production of piezonuclear reactions and neutron emissions, provided
that a given time (and energy) threshold is overcome. Such a
phenomenon is naturally associated to collapse of the imploding
bubbles \cite{carmig7, carmig8}.

It was then hypothesized that the fracture of solid materials would
be able to reproduce the cavitation conditions of liquids and hence
lead to the production of piezonuclear reactions, provided that the
materials were properly selected. In this case, one expects that the
role of cavitation and the induced bubble \ collapse is played by
the compenetration of molecular layers.

Accordingly, tests were conducted to assess neutron production from
piezonuclear reactions in solids subjected to compression till failure \cite%
{carp, carcarp}.

\begin{figure}
\begin{center} \
\includegraphics[width=0.8\textwidth]{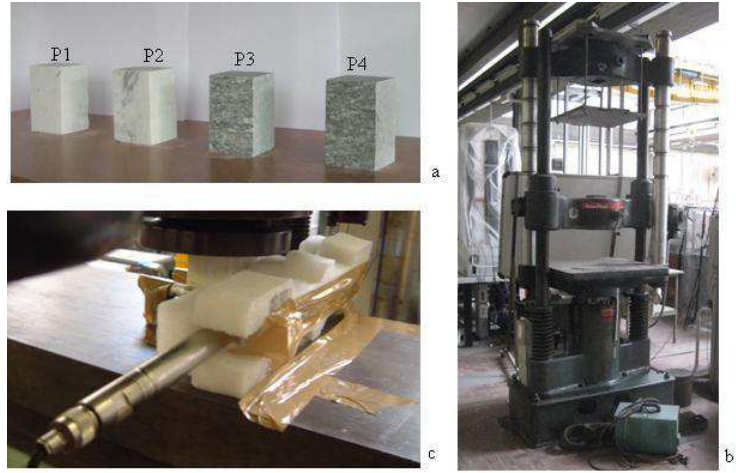} \caption{\textit{The test specimens
analysed, two in Carrara marble (P1, P2) and two in Luserna granite
(P3, P4), measured 6x6x10 }$\mathit{cm}^{\mathit{3}}$\textit{\ (a).
Baldwin servo-controlled press used for the compression tests (b).
Helium-3 neutron detector placed in the proximity of test specimen
P1 during the test. The detector is enclosed in a polystyrene case
for protection against possible impacts due to test specimen failure
(c).}}\label{carp1}
\end{center}
\end{figure}

The materials selected for the tests were Carrara marble (calcite)
and green Luserna granite (gneiss). This choice was prompted by the
consideration that, test specimen dimensions being the same,
different brittleness coefficients \cite{carp2} would cause
catstrophic failure in granite, not in marble. The test specimens
were subjected to uniaxial compression to assess scale effects on
brittleness \cite{carp3}. Four test specimens were used, two made of
Carrara marble, consisting mostly of calcite, and two made of
Luserna granite, all of them measuring 6x6x10 $cm^{3}$
(Fig.\ref{carp1} a). The same machine was used on all the test
specimens: a standard servo-hydraulic press with a maximum capacity
of 500 $kN$, equipped with control electronics (Fig.\ref{carp1} b).
This machine makes it possible to carry out tests in either load
control or displacement control. The tests were performed in piston
travel displacement control by setting, for all the test specimens,
a velocity of 10$^{-6}$ $m/s$ during compression. Neutron emission
measurements were made by means of a helium-3 detector placed at a
distance of 10 $cm$ from the test specimen and enclosed in a
polystyrene case so as to prevent the results from being altered by
acoustical-mechanical stresses (Fig.\ref{carp1} c). During the
preliminary tests, thermodynamic neutron detectors of the bubble
type BD (bubble detector/dosimeter) manufactured by Bubble
Technology Industries (BTI) were used, and the indications obtained
persuaded us to carry on the tests with helium-3 detectors.

\subsection{Experimental set-up}

The servo-controlled press employed works by means of a digital type
electronic control unit. The management software is TestXpertII by
Zwick/Roel, while the mechanical parts are manufactured by Baldwin (Fig.\ref%
{carp1} b). The force applied is determined by measuring the
pressure in the loading cylinder by means of a transducer. The
margin of error in the determination of the force is 1\%, which
makes it a Class 1 mechanical press. The stroke of the press platen
in contact with the test specimen is controlled by means of a wire
type potentiometric displacement transducer.

The neutron detector used in the tests was a helium-3 type with
electronic of preamplification, amplification and discrimination
directly connected to the detector tube, which is of the type
referred to as \textquotedblleft long counter\textquotedblright .
The helium-3 gas supplies the neutron detection signal through the
phenomenon of recoil protons from the anelastic scattering of the
neutrons against the nuclei of the helium-3 atoms; the protons
supply the electric signal which is read by the electronics
equipping the detector. The neutron background was measured at 600
$s$ time intervals to obtain sufficient statistical data with the
detector in the
position shown in Fig.\ref{carp1} c. The average background count rate was $%
\left( 3.8\pm \ 0.2\right) \times 10^{-2}cps$, corresponding to an
equivalent flux of thermal neutrons of $\left( 5.8\pm \ 0.3\right)
\times 10^{-4}nthermal/s\cdot cm^{2}$.

\subsection{The tests}

Neutron emissions were measured on four test specimens, two made of
marble,
denoted with P1, P2, and two of granite, denoted with P3, P4 (see Fig.\ref%
{carp1} a). The test specimens were arranged with the two smaller
surfaces in contact with the press platens , with no coupling
materials in-between, according to the testing modalities known as
\textquotedblleft test by means of rigid platens with
friction\textquotedblright . The mass and density of each marble and
granite test specimen are given in Table~3.
\newline

\begin{tabular}{ccccc}
\multicolumn{5}{c}{Table 3: Physical characteristics of the test
specimens.}
\\
Test specimens & Dimensions [$cm^{3}$] & Material & Weight [$g$] & Density [$%
g/cm^{3}$] \\
P1 & 6x6x10 & Carrara marble & 950 & 2.64 \\
P2 & 6x6x10 & Carrara marble & 946 & 2.62 \\
P3 & 6x6x10 & Luserna granite & 882 & 2.45 \\
P4 & 6x6x10 & Luserna granite & 836 & 2.32%
\end{tabular}
\newline
\newline

\begin{figure}
\begin{center} \
\includegraphics[width=0.8\textwidth]{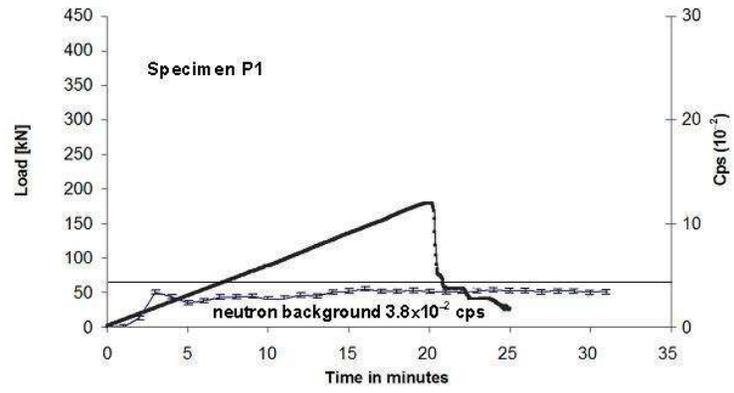} \caption{\textit{Load vs. time and cps
curves for test specimen P1 in Carrara marble.}}\label{P1}
\end{center}
\end{figure}

\begin{figure}
\begin{center} \
\includegraphics[width=0.8\textwidth]{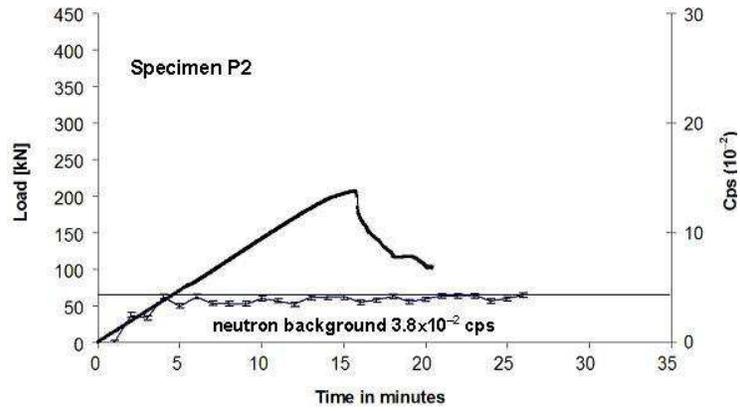} \caption{\textit{Load vs. time and cps
curves for test specimen P2 in Carrara marble.}}\label{P2}
\end{center}
\end{figure}

The electronics of the neutron detector were powered at least 40
minutes before starting the compression tests to make sure that the
behaviour of the device was stable with respect to intrinsic thermal
effects. Then, background measures were repeated for 600 $s$ to make
sure there were no variations. The acquisition time was fixed at 60
$s$ and the results of count rate measurements are given in
Figs.\ref{P1}-\ref{P4} together with the diagrams of the force
applied to the test specimens as a function of the time elapsed
since the beginning of its application.

\bigskip

\subsection{Neutron Emission Measurements}

The measurements of neutron emissions obtained on the marble test
specimens yielded values comparable with the background, even at the
time of test specimen failure. The neutron measurements obtained on
the two granite test specimens, instead, exceeded the background
value by about one order of magnitude at the test specimen failure.

After 20 $min$, test specimen P1 reached a peak load of ca 180 $kN$,
corresponding to an average pressure on the bases of 50 $MPa$; after 15 $min$%
, test specimen P2 reached a peak load of ca 200 $kN$, corresponding
to an average pressure on the bases of 55.6 $MPa$.

Test specimen P3 reached at time T(P3) = 32 $min$ a peak load of ca
$400kN$, corresponding to an average pressure on the bases of 111.1
$MPa$. When failure occurred, the count rate was found to be $\left(
28.3\pm \ 0.2\right) \times 10^{-2}cps$, corresponding to an
equivalent flux of thermal neutrons of $\left( 43.6\pm \ 0.3\right)
\times 10^{-4}nthermal/s\cdot cm^{2}.$

Test specimen P4 reached at time T(P4) = 29 $min$ a peak load of ca 340 $kN$%
, corresponding to an average pressure on the bases of 94.4 $MPa$.
When failure occurred, the count rate was found to be $\left(
27.2\pm \ 0.2\right) \times 10^{-2}cps$, corresponding to an
equivalent flux of thermal neutrons of $\left( 41.9\pm \ 0.3\right)
\times 10^{-4}nthermal/s\cdot cm^{2}.$

\begin{figure}
\begin{center} \
\includegraphics[width=0.8\textwidth]{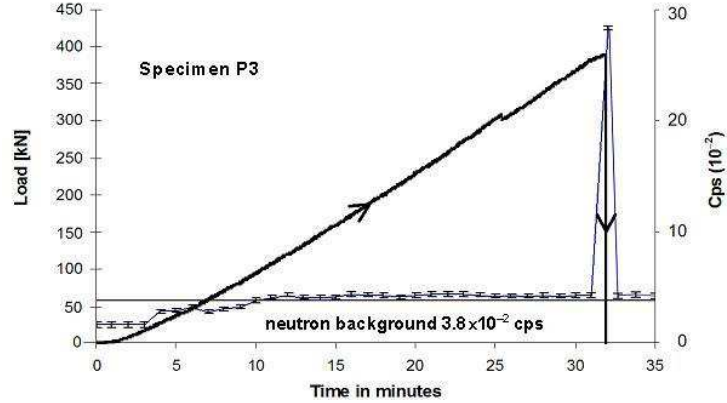} \caption{\textit{Load vs. time and cps
curves for test specimen P3 in Luserna granite.}}\label{P3}
\end{center}
\end{figure}

\bigskip

\begin{figure}
\begin{center} \
\includegraphics[width=0.8\textwidth]{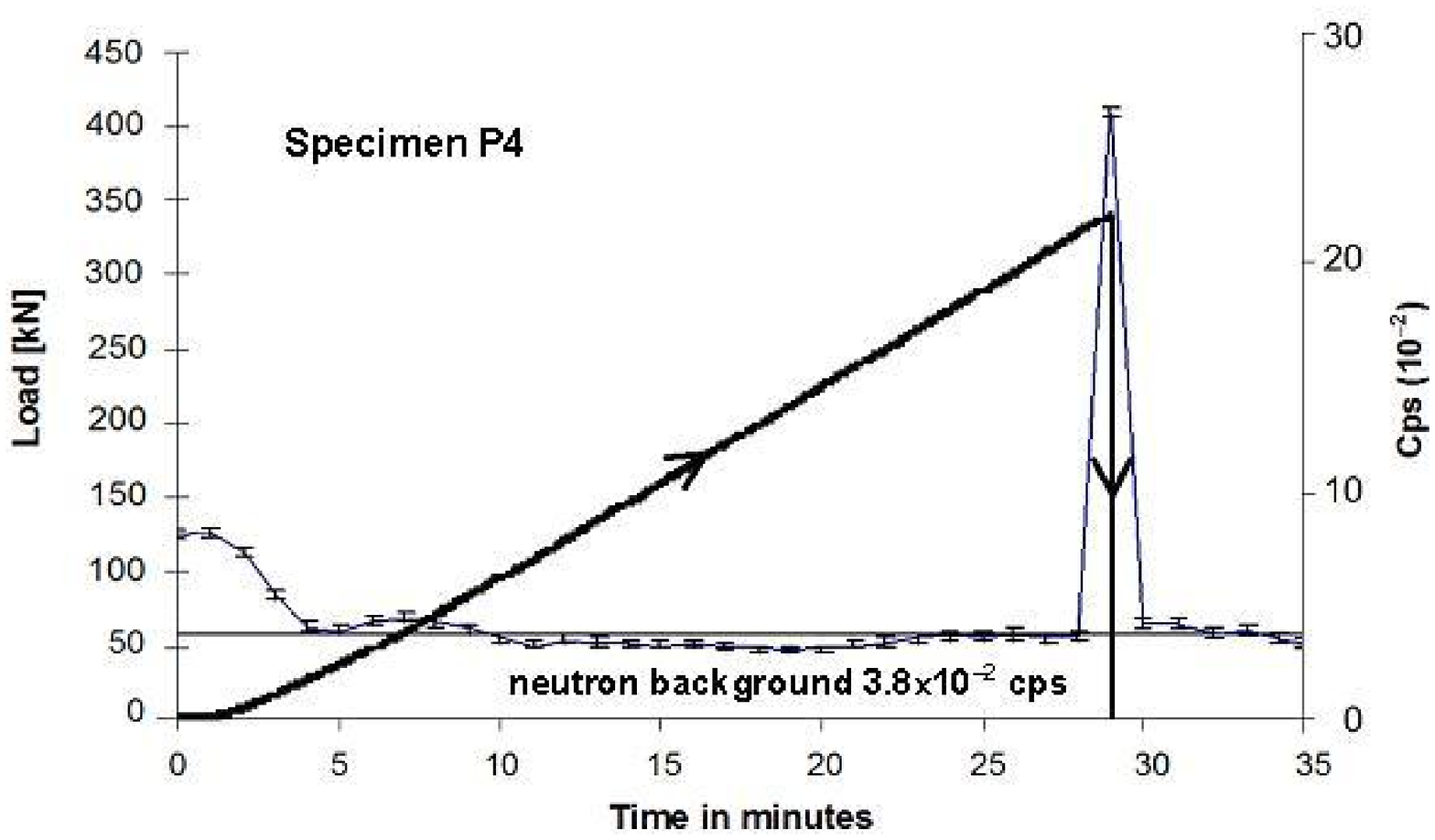} \caption{\textit{Load vs. time and cps
curves for test specimen P4 in Luserna granite.}}\label{P4}
\end{center}
\end{figure}

Notice how the above neutron measurements occurring in P3 and in P4
failure are well beyond the background interval and how the value
obtained on P3 is greater than the value measured on P4. We believe
that this event, albeit with the due caution, may be ascribed to the
unstable failure of these test specimens and the greater quantity of
energy released by P3 compared to P4 at the time of failure. Figures
\ref{P1}-\ref{P4} summarise the evolution of the neutron count rates
together with the load vs. time curves for the four test specimens.

\subsection{Analysis of the Results}

\subsubsection{Factors Involved in Controlling Rock Failure}

Experimental data from rocks tested in compression generally
indicate that this is a brittle material, since it exhibits a rapid
decrease in load carrying capacity when deformed beyond a peak load.
When the softening diagram is very steep, or even shows
simultaneously decreasing strain and stress values, the material is
said to present a snap-back or catastrophic behaviour (cusp
catastrophe). This is in contrast with ductile materials which
retain considerable strength beyond the peak.

Even though in the post-peak phase the load-carrying capacity of the
material may decrease considerably with increasing strain (i.e. the
material \textquotedblleft softens\textquotedblright ), in actual
practice this behaviour can be important to the overall performance
of a rock or concrete structure.

In laboratory studies, the external loading system is the testing
machine. Depending on its design, the machine may have a relatively
soft or stiff characteristic compared to the test specimen. A stiff
frame and electronic servo-controls are required in order to observe
the post-failure behaviour of brittle materials \cite{hud}.

\begin{figure}
\begin{center} \
\includegraphics[width=0.8\textwidth]{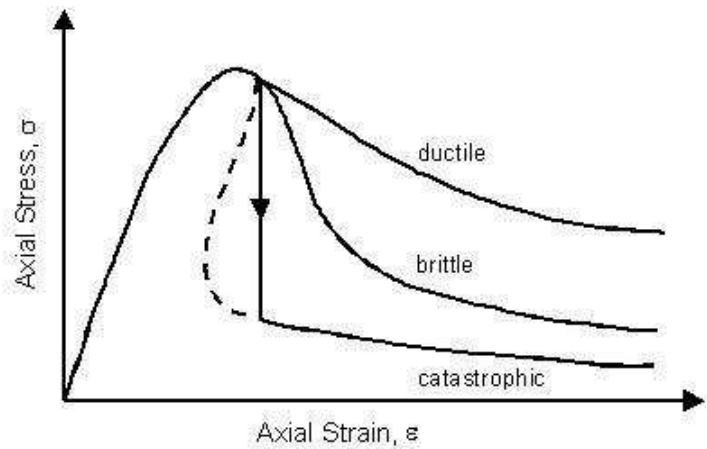} \caption{\textit{Ductile, brittle and
catastrophic behaviour.}}\label{car6}
\end{center}
\end{figure}

By programming a linear increase in axial displacement with time,
the complete stress-strain curve for rock is obtained
(Fig.\ref{car6}). Generally, strains increase linearly with time
even though the specimens fail and the system undergoes a
progressive reduction in load-bearing capacity during the process.
Such specimens are said to exhibit \emph{stable failure}
characteristics.

\begin{figure}
\begin{center} \
\includegraphics[width=0.8\textwidth]{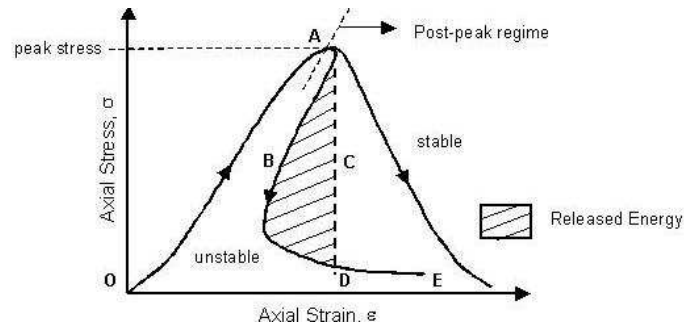} \caption{\textit{Energy release and
stable vs. unstable stress-strain behaviour.}}\label{car7}
\end{center}
\end{figure}

However, if the complete stress-strain curve does not monotonically
increase in strain, the linear increase can only be achieved along
the OACDE curve shown in Fig. \ref{car7}. Since the material
behaviour is represented by the OABDE curve, excess energy,
proportional to the shaded area, is released by the specimen and the
result is uncontrolled violent failure. Such specimens are said to
exhibit \emph{unstable failure} characteristics.

For both traction and compression behaviours, the size-scale
transition from ductile to brittle behaviour is governed by a
nondimensional brittleness number, $s_{E}$, which is a function of
material properties and structure size-scale \cite{carp2,carp3}.

Stable and unstable behaviours, although some materials are
intrinsically
more brittle than others, both depend on the specimen size and shape \cite%
{carp2,carp3,carp4,carp5}. In general, it can be stated that energy
release modalities during compressive tests depend on the intrinsic
brittleness of the material of the test specimens, as well as on
test specimen dimensions and slenderness. Furthermore, it may be
assumed that energy release rate also depends on the velocity
setting of testing machine piston travel.

\subsubsection{Discussion and Remarks}

In this study, all compression tests were conducted through feedback
control of the axial displacement of piston travel on test specimens
having the same dimensions. The complete specimen collapse process
was observed only for P1 and P2 marble specimens, whose behaviour
was seen to be ductile compared to the brittle catastrophic failure
behaviour displayed by granite specimens P3 and P4. For the latter
two, in fact, failure occurred instantaneously, without showing the
descending branch of the load-time curve.

By taking into account only the positive derivative branch of the
load-time curves, from the evolution of the load in the four test
specimens it can be
seen that failure occurs when peak compressive load is reached (Figs. \ref%
{P1}-\ref{P4}). The elastic strain energy accumulated in the test
specimens up to failure, $\Delta E$, is given in Table 4.
\newline
\newline

\begin{tabular}{ccc}
\multicolumn{3}{c}{Table 4: Elastic strain energy at the peak load,
$\Delta
E $ .} \\
Test specimen & Material & $\Delta E(J)$ \\
P1 & Carrara marble & 124 \\
P2 & Carrara marble & 128 \\
P3 & Luserna granite & 384 \\
P4 & Luserna granite & 296%
\end{tabular}
\newline
\newline

As we shall see in the following Section, for each test specimen it
is possible to draw some conclusions on the release rate of the
elastic energy accumulated, considering that granite displays
brittle, catastrophic failure behaviour, while marble is
characterised by a more ductile behaviour. This will allow us to
theoretically explain the different behaviour of specimens P1 and P2
with respect to P3 and P4 as far as neutron emission is concerned.

This experiment clearly shows that piezonuclear reactions giving
rise to neutron emissions are possible in inert non-radioactive
solids, in addition to liquids. In Sect.7\ it will be seen that in
this case, too, the emission
of neutrons is related to a definite behaviour in energy and time \cite%
{carmig7,carmig8}.

Another factor to be taken into account is the composition of the
materials in which piezonuclear reactions may be produced. It should
be noted, in fact, that Carrara marble is basically made of calcite,
i.e. a crystalline form of calcium carbonate CaCO3, whereas in
general granite is made of quartz, alkaline feldspar, plagioclasium
and even biotite and horneblend, of which the last two minerals
contain iron in greater or lesser quantities. The fact that the
marble used in the tests did not contain iron, and granite instead
did, could be another factor contributing to the phenomenon in
question, by analogy with the case of piezonuclear reactions in
liquids. In fact, we have seen in Sect.3\ that piezonuclear
reactions with neutron emissions were obtained in liquids consisting
of iron chloride or iron nitrate subjected to ultrasounds and
cavitation.

\section{Coherence with the Results of other Experiments}

As we have seen in Sects.2,3, our cavitation experiments evidenced
two kinds of phenomena: transformation of nuclides
(\cite{carmig1}-\cite{carmig4}) and neutron emission
(\cite{carcher1,carcher2}). Let us discuss such findings in
connection with the results of other experiments.

As to nuclide production, the findings of such experiments (in
particular of the first one~\cite{carmig1,carmig2}) are similar
under many respects to
those obtained by Russian teams at Kurchatov Institute and at Dubna JINR~%
\cite{uru1}-\cite{uru3} in the experimental study of electric
explosion of titanium foils in liquids. In a first experiment
carried out in water, the Kurchatov group~\cite{uru1,uru2} observed
change in concentrations of chemical elements and the absence of
significant radioactivity. These results have been confirmed at
Dubna~\cite{kuz}. Subsequently, the experiments have been carried
out in a solution of uranyl sulfate in distilled water,
unambiguously showing~\cite{volk} a distortion of the initial
isotopic relationship of uranium and a violation of the secular
equilibrium of $^{234}$Th (as already recalled in Sect.4). Further
experiments are presently being carried out at the Nantes GeM
laboratory, and their preliminary results are in agreement with
those obtained by Urutskoev et al.\cite{priem}. Due to the
similarity of such results with ours, in our opinion the two
observed phenomena have a common origin. Namely, one might argue
that the shock waves caused by the foil explosion in liquids act on
the matter in a way similar to ultrasounds in cavitation. In other
words, the results of the Russian teams support the evidence for
piezonuclear reactions\footnote{%
Another possible interpretation proposed for such phenomena (at
least for the titanium foil explosion) is in terms of the light
magnetic monopoles introduced by Lochak \cite{loch}.}. A
confirmation of this hypothesis comes from the Thorium experiment we
carried out (see Sect.4).

As to neutron emission, we already quoted the Oak Ridge experiment \cite%
{tal1,tal2,tal3,tal4} on possible nuclear fusion in deuterated
acetone subjected to cavitation. The measured neutron flux was said
to be compatible
with d-d fusion during bubble collapse. Some authors disclaimed the results~%
\cite{ss}, others conversely confirmed them~\cite{xb,for}. As to
what the results of our investigations are, one would not be
surprised of the controversial results and hence opinions on the
outcomes of the Oak Ridge experiments~\cite{tal1,tal2,tal3,tal4}.
Our outcomes show that neutron emission is obtained by cavitating
solutions containing Iron and, even if in a very small quantity, by
cavitating pure water. Hence the effects, that we measured, must be
brought about by almost thoroughly unknown mechanisms which are
triggered by pressure. With this in mind, we believe that whoever
tried to reproduce the Oak Ridge experiments must have faced unusual
behaviours and results since along with the very well known and
expected neutrons from D-D fusion, other unknown effects --- like
the existence of a time (energy) threshold for neutron emission ---
would be superimposed, thus generating confused results which do not
precisely confirm the common phenomenological predictions about
fusion.\newline
\qquad Let us stress once again that the experiments~\cite%
{tal1,tal2,tal3,tal4} belong to the research stream known as
sonofusion (or
acoustic inertial confinement fusion). It amounts to the attempt to produce%
\emph{\ known} nuclear reactions by means of ultrasounds and
cavitation. Conversely our case is completely different. We produced
\emph{new} nuclear reactions (piezonuclear reactions) that involve
heavy nuclei but do not, apparently, affect Hydrogen or light ones
(at least within 90 minutes) under unusual conditions, like the
existence of an energy threshold for these reactions to happen and
the apparent lack of gamma emission concomitant to neutron emission.

Our experiments seem also to have some relations with the research
on low
energy nuclear reactions (LENR) or condensed matter nuclear reactions \cite%
{storms}. In many of such experiments, neutron emission was observed
without apparent concomitant emission of $\gamma $ rays. We shall
discuss in more detail \ this point elsewhere\cite{petr}.Let us only
quote for instance the recognition of the important role played by
the CR39 detectors in \bigskip evidencing neutron traces \cite{pam}.

\section{Phenomenological Model of Piezonuclear Reactions}

\bigskip

In this Section, we want to propose a possible mechanism able to
account for piezonuclear reactions, that is constituted by the
integration of two parts. One is essentially classical in nature,
and based on the well-known fact that cavitation allows one to
achieve an extreme concentration of energy for unit time in the
collapsing bubble \cite{flan,chen}. The other, non-classical one
relies instead on a phenomenologically grounded formalism of local
Lorentz invariance breakdown, mathematically based on a deformation
of the Minkowski space-time \cite{carmig6,carmig8}.

\subsection{Classical Cavitation Model}

Let us first consider the case of \ the experiments in cavitating
water.The fundamental and intriguing question to be put is: How can
the pressure waves generated by cavitation trigger nuclear
reactions? The answer comes possibly from the well-known fact that
cavitation allows one to achieve an extreme concentration of energy
for unit time in the collapsing bubble. Indeed there exist
speculations on the possibility that cavitation (in particular,
sonoluminescence) might be a viable approach to inertial-confinement
fusion (provided that the temperature attained in the process is
substantially
higher than that predicted by a simple thermal model, namely 10$^{4}$ $%
^{\circ }K$) \cite{flan,chen}. Let us illustrate a classical model
of piezonuclear reactions, based on the above features of the
cavitation process.

\begin{figure}
\begin{center} \
\includegraphics[width=0.8\textwidth]{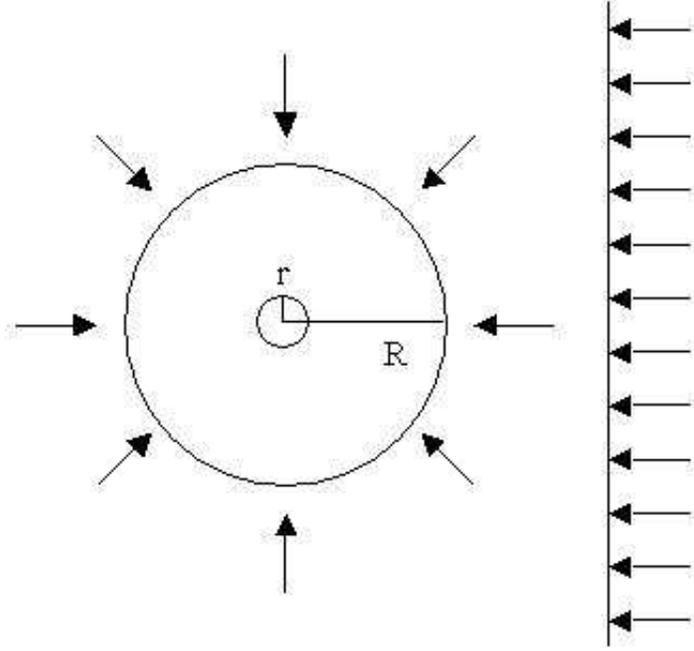} \caption{\textit{Conversion of a plane
pressure wave into a spherical pressure wave on the bubble surface
during the cavitation process.}}\label{conv}
\end{center}
\end{figure}

In order to explain how cavitation can produce the energies needed
to induce nuclear fusion, let us take into consideration the physics
underlying the cavitation process. It consists, as is well known, in
the implosive collapse of a gas bubble within a liquid under
suitable pressure conditions. In our case (water sample) the speed
of sound is about $v$ $\simeq $10$^{3}$ $m/s$, which --- on account
of the used frequency $\nu \simeq $ 10$^{4}$ $Hz$ --- corresponds to
a wavelength $\lambda \simeq $ 10$^{-1}m$. In order to get the gas
bubble to implode, the plane pressure wave of the ultrasounds must
be converted into a symmetric spherical shock wave on the bubble
surface (see Fig.\ref{conv}\textbf{)}.

Therefore, the condition to be satisfied for an implosive collapse
is that the wavelength must be much greater than the bubble size.
Taking as example a spherical bubble with radius $R$, this means
$\lambda >>R$. Since the size of a bubble subjected to cavitation is
of the order of magnitude $R$= 10$^{-6}$ $m$ (see
refs.\cite{carmig7,carmig8}), the collapsing condition is respected
in all the three experiments.

We now remark that --- contrary to what previously believed --- the
only atoms influenced by the shock wave producing the cavitation are
the ones lying on the surface of the bubble itself. These atoms are
trapped by the surface tension of the bubble (generated by the
combined electrostatic repulsion of the liquid and of the bubble) in
a double-layer film at the border liquid-bubble. Namely, all other
atoms inside the bubble volume escape to the outside during the
collapse, due to the fact that the inner pressure, corresponding to
the saturated vapor pressure of water, is far lower than the
external pressure. The trapping of atoms at the bubble surface is
expected to be more effective for metals. Indeed, it is known that
usually metals and metal cations do not enter the cavitation bubble,
but stay at the interface between bubble and bulk solution. At room
temperature (300 $^{\circ }K$) --- as it was the case of all
experiments --- the saturated vapor pressure of water is 0.02 $bar$,
whereas we estimated that the pressure induced by the implosive
shock wave, for a transmitted power of 100 $W$ is of the order of
10$^{9}$ $bar$. This circumstance entails that there is no limit to
the spatial size attainable by the collapse. It is therefore
reasonable to hypothesize that the lower limit of this size can be
identified with the nuclear size. As a matter of fact, there is in
literature no experimental information on the value of the minimum
size attained by a collapsing cavitation bubble. Therefore, we can
suppose that at the end of collapse the bubble dimensions become
near to the nuclear dimensions, commonly about 10$^{-15}m$ (Fermi
radius).

As stated before, if the wavelength $\lambda $ of the plane pressure
wave satisfies the condition $\lambda \gg $ $R$ , where $R$ is the
bubble radius, the plane pressure wave becomes a spherical shock
wave symmetrically acting on the bubble. Let $P$ be the power of the
plane pressure wave, and $R$, $r$ the radius of the bubble before
and after collapse, respectively. Then, the power density on the
bubble before and after collapse are $D_{P}=P$/(4$\pi R^{2}$) and
$D_{P}^{\prime }=P$/(4$\pi r^{2}$). If the initial energy flux (i.e.
the energy for unit time and unit surface) is conserved (due to the
continuity equation and to energy conservation), we have
\begin{equation}
P=SD_{P}=(\pi r^{2})D_{P}^{\prime }\Longrightarrow D_{P}^{\prime
}=D_{P}(S/\pi r^{2})=fD_{P}
\end{equation}%
where $f=$ $(S/\pi r^{2})$ is the amplification factor.

As a consequence, the power density on the bubble surface, generated
by the power of the plane pressure wave, produces after implosion a
power density increased by the factor $f$ on the reduced bubble
surface. Such a factor, caused by the implosion, ranges from $f_{A}$
$\sim $ 10$^{8}$ (when the
bubble collapses to the atomic size, $r_{A}$ = 10$^{-8}$ $cm$) to $f_{N}$ $%
\sim $ 10$^{18}$ (collapse to nuclear radius, $r_{N}$= 10$^{-13}$
$cm$). In the three experiments we used powers ranging from 100 to
630 $W$, i.e. from 6 $\times $10$^{20}$ to 4 $\times $10$^{21}eV/s$.
Thus the final power density $D_{N}$ for collapse to the nuclear
radius is $D_{N}\simeq $2$\times $10$^{46}$ $eV/(s\times cm^{2})$,
corresponding to an equivalent temperature (for a low-density
plasma) of at least 10$^{20}$ $^{\circ }K$. We think that this power
density allows a heavy-ion-fusion-like process between two parent
nuclei to generate a son nucleus. When the bubble collapses and the
bubble surface shrinks to the nuclear dimensions, trapped atoms do
move together with the surface and come closer and closer. The
collapsing surface of the bubble acts therefore as an
\emph{\textquotedblright inertial accelerator\textquotedblright }\
of neutral atoms and the squeezing of the surface to the nuclear
dimension produces on the bubble surface the energy
required to activate the fusion\footnote{%
The fact that the collapsing bubble surface is responsible for the
nuclear reaction ignition is confirmed by the evaluation of the
number of interacting atoms, endowed with the velocity required to
overcome the internuclear Coulomb barrier, present on the bubble
surface for a given overpressure. Indeed, it can be shown that this
number is incompatible with
the number of atoms inside a (even rarefied) bubble (see ref.\cite{carmig7}).%
}. At the final stage of this process, atomic electrons are stripped
away, and a kind of heavy-ion cold fusion (in the sense of
Oganessian \cite{ogan}) can occur.

\subsection{Application to Europium}

Let us apply the model discussed above to the fusion of europium, in
order to explain the results of the third experiment (see
Subsect.2.3). Possible
candidates as parent nuclei for $Eu_{63}^{138}$ are $Zr_{40}^{90}$ and $%
V_{23}^{51}$. A feasible reaction scheme could be:
\begin{equation}
Zr_{40}^{90}+V_{23}^{51}\longrightarrow \ Eu_{63}^{138}+3n.
\end{equation}

Where the nuclides $Zr$ and $V$ could come from? A possible answer
to this question can be found in the impurities lying on the surface
of the sonotrode tip. The latter was shaped through mechanical tools
(lathes) made by alloys of iron, vanadium and zirconium, introduced
to harden the tools themselves. During the manufacturing, small
numbers of atoms should remain trapped inside the iron lattice of
the sonotrode tip. Impurity atoms are more loosely bond to the iron
lattice than the iron atoms, so the ultrasonic vibrations of the tip
can remove them from the lattice. By the way, the possibility of the
neutron excess (see refs.\cite{carmig1}-\cite{carmig4}) could be
explained by the observations already made by other research groups
working on cavitation-induced nuclear reactions
\cite{tal1}-\cite{kriv}.

The Coulomb barrier against fusion for $Zr$ ($Z_{1}$=40, $A_{1}$=90)
and $V$ ($Z_{2}$=23, $A_{2}$=51) can be evaluated by the formula
\begin{equation}
E_{coul}=\frac{Z_{1}Z_{2}}{A_{1}^{1/3}+A_{2}^{1/3}}MeV=112MeV
\end{equation}
or also by
\begin{equation}
E_{coul}=\frac{Z_{1}Z_{2}}{d}MeV=140MeV,
\end{equation}
with $d=r_{1}+r_{2}+2r_{0}$ (where $r_{1}$=4.5 $\times $10$^{-13}$ $cm$ and $%
r_{2}$=3.7 $\times $10$^{-13}$ $cm$ are the nuclear radii of $Zr$
and $V$, respectively, and $r_{0}$ = 0.5 $\times $10$^{-13}$ $cm$ is
the characteristic Bohr-Wheeler nuclear length).

From the power density $D_{PN}$ on the bubble surface after
collapsing from a radius of 10$^{-4}$ $cm$ to the nuclear radius
$r_{N}$ estimated above, it is possible to evaluate the energy and
the energy per nucleon needed to bring about the formation of
europium 138 from vanadium and zirconium according to reaction (6).

\begin{figure}
\begin{center} \
\includegraphics[width=0.8\textwidth]{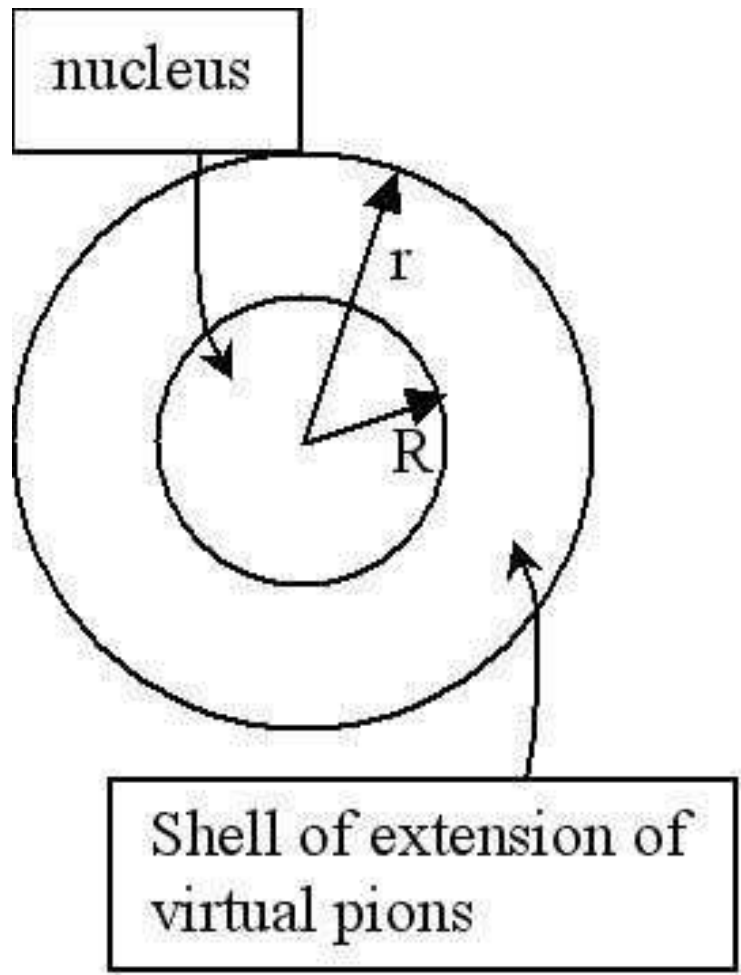} \caption{\textit{Nuclear radius and
shell extension of the virtual pions around the
nucleus.}}\label{pions}
\end{center}
\end{figure}

With reference to Fig.\ref{pions}, let $r_{N}$ and $R_{N}$ be,
respectively, the radius of the nucleus and the radius of the shell
of extension of the virtual pions. One has
\begin{eqnarray}
r_{N} &=&0.5\times 10^{-13}\sqrt[3]{A}\text{ }cm;  \notag \\
R_{N} &\simeq &R_{F}=0.5\times 10^{-13}cm.
\end{eqnarray}%
Then, the effective nuclear radius $\overline{R}$ is given by
\begin{eqnarray}
\overline{R} &=&\frac{r_{N}+R_{N}}{2}=\frac{1}{2}\left( \frac{1}{2}\sqrt[3]{A%
}+\frac{1}{2}\right) 10^{-13}cm=  \notag \\
&=&0.25\times \left( \sqrt[3]{A}+1\right) 10^{-13}cm\simeq r_{N}.
\end{eqnarray}%
The effective nuclear surface $S$ is therefore $S=4\pi r_{N}^{2}=A^{\frac{2}{%
3}}\pi 10^{-26}cm^{2}$. For vanadium and zirconium one gets,
respectively:
\begin{eqnarray}
\text{For vanadium}\text{: } &&S_{V}=4.3\times 10^{-25}cm^{2};  \notag \\
\text{For zirconium}\text{: } &&S_{Z}=6.3\times 10^{-25}cm^{2}.
\end{eqnarray}%
The mean value is $\overline{S}\simeq 5.3\times 10^{-25}cm^{2}$.

In order to find the energy achieved in this case, the collapse time
of the bubble must be estimated. For an employed frequency $\nu $=20
$kHz$, the period is $T=1/\nu $= 5$\times $10$^{-5}s$. The collapse
time is obviously comprised between $T$ (upper limit) and the time
$t_{s}=R/v_{s}$ taken by
the pressure wave to travel the bubble radius (at the sound speed): $%
t_{s}<t_{c}<T$. In deionized and bidistilled water, the sound speed is $%
v_{s}\simeq $ 1.4$\times $10$^{3}$ $m/s$. This corresponds to the
lower time limit for collapse $t_{s}$ = 7$\times $10$^{-10}s$. The
energy transmitted by the bubble implosion to the $V-Zr$ system is
therefore
\begin{equation}
E_{c\min }=\overline{S}D_{PN}t_{s}<E_{c}<E_{c\max
}=\overline{S}D_{PN}T
\end{equation}

One gets
\begin{eqnarray}
E_{c\min } &=&3.3\times 10^{6}MeV;  \notag \\
E_{c\max } &=&2.5\times 10^{11}MeV.
\end{eqnarray}
The total number of nucleons is $N$=90+51=141. Thus the energy for nucleon $%
\varepsilon =E/N$ ranges in the interval
\begin{equation}
\varepsilon _{\min }\simeq 2.3\times 10^{4}MeV<\varepsilon
<\varepsilon _{\max }\simeq 1.8\times 10^{9}MeV.
\end{equation}

Let us notice that $\varepsilon _{\min }$ is much higher than the
maximum estimated Coulomb barrier for the europium formation.
Moreover, the range of values of $\varepsilon $ corresponds to the
energies of the direct nuclear reactions induced by photons and to
the production of pairs $e^{+}e^{-}$ due to photon conversion in the
nuclear electric field, many orders of magnitude higher than the
energy required to overcome the Coulomb barrier.

Then, one can conclude that the mechanism proposed for piezonuclear
fusion is able indeed to account for europium formation generated by
the cavitation process.

The simple phenomenological model discussed above may of course be
also responsible of the results we got in the first two experiments
(see Sect.2). In this case, the atoms of some elements (the
decreasing ones of the first experiment) may have been subjected to
the inertial-fusion-like process due to bubble collapse to generate
new elements (the increasing ones and/or the transuranic elements
observed in the second experiment).

Moreover, this mechanism avoids the need for the introduction of a
nuclear shape deformation, invoked sometimes to increase the nuclear
tunnelling probability. We remind, to this regard, that $Zr$ and $V$
are typically spherically shaped nuclei, at least in their
fundamental state, so that a shape deformation would be very
difficult to justify.

On the other hand, an Oganessian-like low temperature heavy ion
nuclear fusion \cite{ogan} could be possible even by the following
mechanism. Let us suppose that there be some kind of oversaturated
vapor of the $\pi $-meson (boson) gas, caused by the final power
density, all around the parent nuclei. The $\pi $-mesons are
emitted, but they are absorbed more slowly --- as if they would
condensate (something like a Bose-Einstein condensation)
---, typically after a time which, on the basis of the previous values of
the energy-space-time density, we estimate as 10$^{-16}$ $s$ (to be
compared with the typical nuclear time of 10$^{-23}$ sec.). As a
consequence, the probability of nuclear interactions is enhanced and
there is an anomalous increase of the nucleus-nucleus cross section.

\subsection{Limits of the Classical Model:the Issue of Spontaneous Fission}

In spite of its effectiveness in accounting \textit{e.g.} for the
europium production, the classical model illustrated above suffers
from some drawbacks we are going to discuss.

First of all, it must be noticed that not all of the power from the
sonotrode is concentrated in one bubble of nuclear dimension. Also,
much of the power put in is dissipated in other processes (in fact,
energy is subtracted \textit{e.g.} by sonoluminescence and by
endothermic sonochemical reactions. This could be taken into account
by an approximated energy efficiency factor, useful for
non-sonochemical reactions; the latter are to be identified with
those reactions that involve parent nuclei to generate composite
nuclei with a greater mass. Moreover, a crucial parameter is just
represented by the final size reached by the implosed bubble, which
critically determines the energy available for the fusion (or
fission) process.

Another intrinsic limit of this model is due to the very
phenomenology of nuclear fusion of heavy nuclei. At energies below
the barrier, the fusion probability is low and rises exponentially
with energy. At energies just above the barrier, the probability for
parent nuclei to fuse and form a composite nucleus does increase
further but not indefinitely. As a matter of fact, at energies of
about 10 $MeV$ per nucleon, the cross section for forming a
composite system is low and diminishes at a rate of 1/$E_{cm}$
(where $E_{cm}$ is the center-of-mass energy of the relative motion
of the two nuclei). Such a decrease occurs because the composite
system (for instance, $V+Zr$), formed at such high relative
energies, is a highly excited nucleus and has large angular
momentum. This causes the composite
system to fission instantaneously (with a time frame much shorter than 10$%
^{-22}$ $s$). For nuclear collisions at the energies we estimated,
the fusion does not occur at all.

\subsection{Deformed Spacetime of Strong Interaction}

A possible answer to both these questions may come from the
formalism of Deformed Special Relativity (DSR). It is a
generalization of Special Relativity (SR) based on a
\textquotedblright deformation\textquotedblright\ of Minkowski
space-time, assumed to be endowed with a metric whose
coefficients depend on the energy of the process considered \cite%
{carmig6,carmig8}. The DSR formalism provides a geometrical
description of the four fundamental interactions (electromagnetic,
weak, strong and gravitational), in terms of phenomenological
deformed metrics with a threshold behaviour. In particular, the
nuclear strong metric reads

\begin{gather}
g_{DSR,strong}(E)=  \notag \\
=diag\left(
b_{strong}^{2}(E),-b_{1,strong}^{2}(E),-b_{2,strong}^{2}(E),-b_{strong}^{2}(E)\right)
;
\end{gather}%
\begin{gather}
b_{strong}^{2}(E)=\left\{
\begin{array}{ll}
1, & 0\leq E<E_{0strong} \\
(E/E_{0strong})^{2}, & E_{0strong}<E%
\end{array}%
\right. =  \notag \\
=\smallskip 1+\Theta (E-E_{0,strong})\left[ \left( \frac{E}{E_{0,strong}}%
\right) ^{2}-1\right] ,  \notag \\
E>0;
\end{gather}%
\begin{equation}
b_{1,strong}^{2}(E)=\left( \sqrt{2}/5\right) ^{2};
\end{equation}%
\begin{equation}
b_{2,strong}^{2}=(2/5)^{2},
\end{equation}%
with

\begin{equation}
E_{0,strong}=\left( 367.5\pm 0.4\right) GeV.
\end{equation}%
where $\Theta (x)$ is the Heaviside theta function, and
$E_{0,strong}$ is the energy value at which the metric parameters
are constant, i.e. the metric becomes Minkowskian.

Let us explicitly notice that the hadronic metric (15)-(19) is not
always isochronous with the usual Minkowski metric $(b_{0}^{2}=1)$.
Actually, it follows from (16) that it is $b_{0}^{2}\not=1$ for
$E_{0,strong}<E$.

Such a case is not new; indeed, as is well known, the same happens
for the gravitational interaction, as shown \textit{e.g.} by the
various measurements of red or blue shifts of electromagnetic
radiation in a gravitational field, or by the relative delays of
atomic clocks put at different heights in presence of gravity.

Let us investigate the possible implications of such an
anisochronism of the hadronic metric. We denote by $dt_{had}$ the
time interval taken by a certain hadronic process for a particle at
rest (\textquotedblright hadronic clock\textquotedblright ) The same
process, when referred to a Minkowskian electromagnetic metric, will
take a time $dt_{e.m.}$ to happen. The former time corresponds to
the real time felt by the particle in its local frame, namely to the
proper time $\tau _{DSR}$ in the Minkowski space deformed by the
strong interaction. The latter is the coordinate time of an external
observer, who looks at the process by means of his electromagnetic
instruments\footnote{%
The interpretation of $dt_{e.m.}$ as the coordinate (Minkowskian)
time is supported also by the fact that the energy of an hadronic
process is in general much higher than the electromagnetic energy
threshold $E_{0,e.m.}$, and therefore the e.m. metric is fully
Minkowskian.}. The two times are
related by the equation%
\begin{equation}
d\tau _{DSR}=b_{0}dt_{0}
\end{equation}%
Such relations is analogous to that between proper time and
coordinate time
found in General Relativity ($d\tau _{GR}=\sqrt{g_{00}}dt$)\footnote{%
Indeed, since $g_{DSR00}=b_{0}^{2}$, the general-relativistic relation for $%
\tau $ becomes exactly the DSR relation (20).}. In the DSR case,
too, the proper time \emph{does not coincide} with the time measured
by the local observer in the particle frame. Like in GR, therefore,
one has to distinguish between the \emph{real (proper) time }$\tau $
and the coordinate (or \emph{universe}) time $t$. As is well known,
such a distinction is fundamental, within GR, for the analysis of
gravitational phenomena (like gravitational collapse). The situation
reminds that of the gravitational fall of a particle toward a
collapsing body. In that case, the proper time is the real time
measured by the particle (influenced by the body gravitational
field), whereas the coordinate time is that measured by a
distant observer in fully Minkowskian conditions. Therefore, $dt_{had}$ and $%
dt_{e.m.}$ are related by
\begin{equation}
{\frac{dt_{had}}{dt_{e.m.}}}=b_{0,strong}
\end{equation}%
or, on account of the explicit form the strong metric (16):
\begin{equation}
{\frac{dt_{had}}{dt_{e.m.}}}=\left\{
\begin{array}{cc}
1, & 0\leq E\leq E_{0,strong} \\
E_{0,strong}/E, & E_{0,strong}\leq E%
\end{array}%
\right. .
\end{equation}

Eq.(22) provides\emph{\ the law of time deformation in a hadronic
field}. It is easily seen that there is isochronism at low energies.

Let us recall that the energy $E_{c}$ attained for collapse of a
cavitating bubble to the nuclear radius ranges in the interval
between $E_{cmin}\sim $ 10$^{6}MeV$ \ and $E_{cmax}\sim $ 10$^{11}$
$MeV$. Then, $E_{cmin}$ is about one order of magnitude higher than
the threshold energy of the DSR strong interaction,
$E_{0,strong}\simeq $ 4$\times $10$^{5}$ $MeV$. This entails that
\emph{the hadronic interaction between nuclei occurs in the
non-Minkowskian part of the strong interaction metric, i.e. the
(hadronic) space-time geometry is deformed in the final collapse
region}. As already noted, in the (usual) flat Minkowskian metric
the interacting nuclei, although overcoming the Coulomb barrier,
produce, after fusion, a nucleus in an highly excited state, and
therefore with high probability of spontaneous fission. On the
contrary, in the deformed spacetime produced by the over-threshold
hadronic condition, the excess energy after fusion goes in deforming
the spacetime region, thus leaving the son nucleus (produced by the
fusion of the two parent nuclei) in a low-excited (or even
unexcited) state.

Therefore,\emph{\ the stability of the nuclei produced by cavitation is due }%
(according to DSR)\emph{\ to the deformation of spacetime in the
collapse region ensuing from the non-Minkowskian behavior of the
strong interaction
in the range }$\emph{E}_{\emph{c}}$\emph{\ $>$ E}$_{0,strong}$%
\emph{. }This explains why no emission of radiation was observed in
the cavitation experiments.

Let us also stress that the presence of the hadronic time deformation \emph{%
contributes also to the lowering of the Coulomb barrier.} In fact,
as is well known - as firstly realized by Levi-Civita - in a
gravitational field the time coefficient $g_{00}$ of the metric acts
as a refractive index, so
the intensity of an electric field $\mathcal{E}$ is changed \ into $\mathcal{%
E}/\sqrt{g_{00}}$. The same happens in the case of a deformed
spacetime, in
which the \ Coulomb barrier becomes $\mathcal{E}_{coul}/\sqrt{b_{0strong}^{2}%
}.$This further enhances the probability of fusion processes (like
in the case of Europium).

\subsection{Threshold Energy for Piezonuclear Reactions}

Let us show that DSR is also able to predict the cavitation power
needed to produce piezonuclear reactions in a stable way. This is a
consequence of the law of time deformation in an hadronic field,
Eq.(22), we rewrite here for reader's convenience:
\begin{equation}
\frac{dt_{hadr.}}{dt_{e.m.}}=\frac{E_{0,strong}}{E}.
\end{equation}

A way to read Eq.(23) is as follows. It can be regarded as an
action-reaction relation, i.e. as the equality between two energy
speeds: An electromagnetic speed $W_{e.m.}$ of supplying energy to
the atoms by the electromagnetic interaction (action) and an
hadronic speed $W_{strong}$ of response by the strong interaction of
nuclei (reaction)
\begin{equation}
W_{strong}=\frac{E_{0,strong}}{dt_{hadr.}}=\frac{E}{dt_{e.m.}}=W_{e.m.}.
\end{equation}

In order to attain the threshold of LLI breakdown for strong nuclear
interaction, during the time taken by a generic cavitating bubble to
collapse, for a given electric energy $E$, an energy speed
$W_{e.m.}$ must be supplied such to equate the nuclear one.

Let $dt_{hadr.}$ the nuclear reaction time given by
\begin{equation}
dt_{hadr.}=\gamma _{strong}\Delta t
\end{equation}%
where $\gamma _{strong}$is the deformed strong relativistic factor and $%
\Delta t=h/m_{\pi }$ is the Yukawa time (nuclear year).

An estimate of $dt_{hadr.}$ at the energy threshold $E_{0,strong}$
can be gotten by means of the relation $\gamma
_{strong}=E_{0,strong}/m_{\pi }$ (on account of the well-known fact
that $\gamma =E/m$ is the relativistic factor of time dilation in
Minkowskian conditions for $E\leq E_{0,strong}$, and by recalling
that the process occurs approaching $E_{0,strong}$ from below).
Replacing such an expression of $\gamma _{strong}$ in Eq.(25) yields
\begin{equation}
dt_{hadr.}=\frac{h}{m_{\pi }^{2}}E_{0,strong}
\end{equation}%
($h=$4.136$\times $10$^{-15}eV\times s$ ; $m_{\pi }=(m_{\pi }^{\pm
}+m_{\pi }^{0})/2=$1.373$\times $108$eV$).

For the energy of the electric interaction we have
\begin{equation}
E=dt_{e.m.}E_{0,strong}\frac{m_{\pi }^{2}}{E_{0,strong}h}%
=dt_{e.m.}W_{strong}.
\end{equation}%
Since $E=dt_{e.m.}(m_{\pi }^{2}/h)$, $W_{strong}$ reads
\begin{equation}
W_{strong}=m_{\pi }^{2}/h=4.8\times 10^{30}eV\times s^{-1}=7.6\times
10^{11}W.
\end{equation}

Let us assume for $dt_{e.m.}$ the time taken by a microbubble of
radius $R$ to collapse to the nuclear size with $r$ $\sim $
10$^{-13}cm$ (due to the electric repulsion of the water atoms
subjected to the ultrasonic pressure
wave). The collapse can occur at the velocity of sound in distilled water, $%
v=v_{s}=$1.4$\times $10$^{3}m/s$, or at the velocity of the shock wave, $%
v=v_{u}=4v_{s}$. Because the ultrasound wavelength is much greater
than the microbubble diameter, it is $dt_{e.m.}=R/v$ in either case.

Therefore we have, for the threshold energy $E_{thres}$:
\begin{equation}
E_{thres}=\frac{Rm_{\pi }^{2}}{vh}.
\end{equation}

The values of $E_{thres}$ deduced from Eq.(29) range from 5$\times
$10$^{2}J$ to 2$\times $10$^{3}J$ for the collapse speed $v_{s}$
(with radius $R$ of
the collapsing microbubbles varying from 1 $\mu $ to 4 $\mu $) and from 10$%
^{2}J$ to 2$\times $10$^{3}J$ for $v_{u}$ (with 1 $\mu $ $<$ $R<$ 8
$\mu $).

In order to produce stable piezonuclear reactions, and therefore a
stable emission of nuclear radiation, it is necessary to supply
constantly an energy $E\geq E_{thres}$ to the system of distilled
water and solute. Such a condition permits to trigger piezonuclear
reactions in presence of broken local Lorentz invariance.

By using a cavitator absorbing 2000 $W$ and able to provide a stable
supply from 100 $J$ up to a maximum of 2000 $J$, it is possible to
investigate the collapse of bubbles with size ranging from 1 $\mu $
to 8 $\mu $, by taking either $v_{s}$ or $v_{u}=$4$v_{s}$ as
collapsing speed.

As we have seen, the existence of the threshold $E_{thres}$ is a
direct consequence of the existence of of the energy threshold
$E_{0,strong}$ for the hadronic interaction.

If the interaction among nuclei occurs in non-Minkowskian conditions (for $%
E> $ $E_{0,strong}$), the excess energy is partly absorbed by the
hadronic
space-time deformation, so that \emph{there is no emission of }$\gamma $%
\emph{\ radiation.}

\emph{The two facts of the energy threshold overcoming, }$E>$\emph{\ }$%
E_{0,strong}$\emph{\ , and of the neutron emission in absence of }$\gamma $%
\emph{\ radiation do provide the complete signature of piezonuclear
reactions produced by the cavitating collapse of gas bubbles of
water in non-Minkowskian conditions. }Both these conditions are met
in the experiments we described.

The existence of an energy threshold to be overcome explains why
iron - which is the most stable element from the point of view of
standard nuclear reactions - is instead favoured in order to get
piezonuclear processes. Indeed, since iron has the highest value of
the binding energy for nucleon, it is the first to attain the energy
threshold needed to deform spacetime,
power supplied being equal\footnote{%
This remark is due to W. Perconti.} \ .

However, it must be stressed that there is a third condition to be
satisfied in order that piezonuclear reactions do occur. This is the
overcoming a power threshold, and was evidenced in the experiments
of solids subjected to mechanical pressure. We shall discuss this
point in Subsect.7.7.

\subsection{Piezonuclear Thorium Decay in DSR}

Let us discuss now the implications\ of DSR for the thorium decay.
The spontaneous decay of $Th^{228}$ through the weak interaction
halves it in a time $t_{1/2}$= 1.9 $years$ = 9.99$\times 10^{5}$
$min$. The ratio between
the half life of thorium, $t_{1/2}$ , and the time interval of cavitation, $%
t_{c}$ = 90 $min$, is $t_{1/2}/t_{c}$ = 10$^{4}$. This means that
cavitation brought about the transformation of $Th^{228}$ at a rate
10$^{4}$ times faster than the natural leptonic decay would do. On
the other hand, the experiments of the Russian team provided no
evidence of spontaneous fission. One is therefore led to deem that
we are facing a phenomenon of accelerated
transformation of thorium into some other nuclide (maybe Lead)\footnote{%
Private Communications by L. Urutskoev.}, induced by cavitation, is
rather due to strong interaction, in particular to its
non-Minkowskian part.

By bearing this in mind, it is possible to interpret $t_{1/2}/t_{c}$ = 10$%
^{4}$ as the ratio between the decaying time of $Th^{228}$ via the
leptonic
interaction (leptonic time $t_{lep}$), and the transformation time of $%
Th^{228}$ via the hadronic interaction (hadronic time $t_{had}$).
Namely, one has:
\begin{equation}
\frac{t_{1/2}}{t_{c}}=10^{4}=\frac{t_{lep}}{t_{had}}.
\end{equation}

Let us recall that the time coefficients of both metrics of
electromagnetic and weak interactions, $b_{0,e.m.}$ and
$b_{0,weak}$, are equal to each other, energy independent, and
always equal to 1. Either metrics is therefore always Minkowskian in
time. Moreover, the space coefficients of both metrics have the same
energy behavior. Thank to this circumstance, it will be always true,
for the intervals of time $dt_{e.m.}$ and $dt_{weak}$, that
$dt_{e.m.}$=$dt_{weak}$. Hence we can write:
\begin{equation}
\frac{t_{weak}}{t_{had}}=\frac{t_{e.m.}}{t_{had}}=10^{4}.
\end{equation}

On account of the hadronic law of time deformation (23), the same
relation of proportionality holds between the threshold hadronic
energy and the hadronic time and between the electromagnetic energy
and the electromagnetic time. In the present case, Eq.(31) can be
rewritten in terms of the time intervals $t_{weak}$ = $t_{e.m.}$ and
$t_{had}$ as:
\begin{equation}
\frac{t_{had}}{t_{weak}}=\frac{E_{0,had}}{E_{e.m.}}.
\end{equation}

From the above relation it is possible to estimate the unknown variable $%
E_{e.m.}$, \textit{i.e. }the energy transferred by the electrical
(Minkowskian) interaction to the nuclei of thorium (which get
transformed into other nuclides by the strong interaction). One gets
\begin{equation}
E_{e.m.}=E_{0,had}\frac{t_{e.m.}}{t_{had}}=367.5GeV\times
10^{4}=3.675\times 10^{15}eV.
\end{equation}

This value of energy is compatible with the maximum energy for nucleon $%
\varepsilon _{\max }$ estimated for the cavitation experiment which
provided evidence of the production of the europium isotope 138.

\subsection{DSR and Piezonuclear Reactions in Solids}

We want to see that DSR allows one to understand the results
concerning the emission of neutrons from brittle fracture of solids
subjected to mechanical pressure. One of the conditions to be met
for piezonuclear reactions to take
place is that the ratio, $r$, between the power of released energy, $%
W=\Delta E/\Delta t$, and the power threshold%
\begin{equation}
Wstrong=7.69\times 10^{11}W=4.8\times 10^{30}eV/s
\end{equation}%
be greater than or equal to 1:%
\begin{equation}
r=W/Wstrong\geq 1.
\end{equation}

Accordingly, based on the data obtained from the tests, the time
interval of released energy, $\Delta t$, in granite test specimens
in which piezonuclear
reactions have occurred, should satisfy the following relationship:%
\begin{equation}
\Delta E/\Delta t\geq Wstrong,
\end{equation}%
and hence:%
\begin{equation}
t\leq \frac{\Delta E}{\Delta Wstrong}=\frac{384}{7.69\times 10^{11}}%
=0.5\times 10^{-9}s=0.5ns.
\end{equation}

Equation (36) was written by considering the energy accumulated in
P3 which was greater than the energy accumulated in P4. For the
marble test specimens, in which peak load is clearly followed by a
softening branch, energy release surely occurred over a period of
time too long to permit the production of piezonuclear reactions.
Accordingly, neutron emissions in granite may be accounted for by
the fact that the power threshold for piezonuclear reactions is
exceeded, as well as by the type of catastrophic failure that
occurs, which entails a very fast energy release, over a time period
of the order of a nanosecond. Furthermore, with these assumptions,
energy release time being the same, it is possible, albeit with the
necessary caution, to ascribe the greater neutron emission from P3
compared to P4 to the fact that $r$(P3) $>$ $r$(P4).

Considering that the elastic strain energy accumulated in specimen
P3 is released at the pressure wave velocity $v$ (for granite $v\sim
4000$ the
extension of the energy release zone results to be equal to:%
\begin{equation*}
\Delta x=vt\sim 4000m/s\times 0.5ns\sim 2%
\mu
m.
\end{equation*}

Such energy release band width $\Delta $x could correspond to the
critical value of the interpenetration length $w_{cr}^{c}$ assumed
by Carpinteri and Corrado \cite{carp5} to explain the critical
conditions for the catastrophic behaviour of solids in compression.
This supports the assumption we made on the role (similar to that of
cavitation in liquids) played in solids by the interpenetration of
molecular layers. In fact, if one considers the elastic energy
stored in the sample and the temperature equivalent to this energy
together with the final pressure before fracture, the material of
the sample is in a region of the phase-space corresponding to a
transition from solid to liquid phase. Our conjecture is that in the
interpenetration layer of thickness $w_{cr}^{c}$ the conditions are
realized for a high density fluid,over-saturated but in metastable
conditions. On the other hand,these locally extreme conditions could
catalyse in the interpenetration band the formation of a plasma from
the gases which are present anyway in the solid materials even at
room conditions.

Anyhow, it is also evident that the availability of an amount of
stored energy for the reactions exceeding the microscopic nuclear
deformed space-time threshold $E_{0,strong}=5.888\times
10^{-8}J=3.675\times 10^{11}eV $ is not sufficient \textit{per se}.
The energy must be\ contained in a space and time (and energy)
hypervolume such that $r\geqslant 1$, i.e., such that the phenomenon
will actually develop in deformed space-time conditions
\cite{carmig7,carmig8}. From Table 4, in fact, it can be seen that
it was $\Delta E>E_{0},strong$ in all the test specimens loaded in
compression, but $r$ was greater than 1 only in granite test
specimens. Hence, even for piezonuclear reactions in solids, the
notion of stored energy must be combined with the notion of speed of
energy release as is the case for liquids.

The fact that the marble used in the tests contains only iron
impurities (not more than 0.07\% of $Fe_{2}O_{3}$), and granite
instead contains a considerable amount of iron (around 3\% of
$Fe_{2}O_{3}$), could be another factor contributing to the
phenomenon in question, by analogy with the case of piezonuclear
reactions in liquids. In fact, piezonuclear reactions with neutron
emissions were obtained in liquids containing ironchloride or iron
nitrate subjected to ultrasounds and cavitation \ (see Sect.3).
Moreover, we have stressed at the end of Sect.3 \ that in the
experiments on liquid solutions, aluminum atoms appeared at the end
in a final quantity as large as about seven times the small initial
quantity.\ This can be explained by assuming that iron undergoes the
piezonuclear fission reaction Eq.(4). This should have occurred also
in the compression tests on granite specimens. Such a conjecture is
supported also by the following considerations of a very general
nature. The present natural abundance of aluminum (7--8\% in the
Earth crust), which is less favoured than iron from a nuclear point
of view (it has a lower bond energy per nucleon), is possibly due to
the piezonuclear fission reaction (4). This reaction --- less
infrequent than one could think --- would be activated where the
environment conditions (pressure and temperature) are particularly
severe. If we consider the evolution of the percentages of the most
abundant elements in the Earth crust during the last 3 billion
years, we realize that iron and nickel have drastically diminished,
whereas aluminum, silicon and magnesium have as much increased. It
is also interesting to realize that such increases have developed
mainly in the tectonic regions, where frictional phenomena between
the continental plates occurred \cite{carp,carcarp}.

\section{Perspectives}

In conclusion, let us briefly outline possible developments and
perspectives of the research on piezonuclear reactions.

Due to the emission of neutrons in bursts, as reported above,
neutron measurements during piezonuclear reactions have very
peculiar features, which make them different from common
measurements of neutrons from neutron emitting elements or nuclear
reactors. We stated that passive detectors are more suited to make
out piezonuclear neutrons than active ones, although the latter were
able to detect them as well. From this perspective it would be
interesting and useful to perform neutron measurements by different
techniques either passive or active in order to collect as more
evidences and features of these emissions as possible. It would be
possible to use elements in foils like Au-197, Mn-55 or Eu-151 and
measure their activation as a passive technique or it would also
possible to use fission chambers as an active one.

As to cavitated liquids, a systematic investigation of solutions of
different salts would be needed, in order to check if iron does
indeed\ play a privileged role.

New experiments can be envisaged in order to exert pressure on
solids and attempt to achieve further evidences of piezonuclear
reactions from solid materials.

In particular, it will be possible to apply ultrasounds to solids,
beyond liquids, like bars of different materials containing or not
containing iron, or bars of iron or steel.

Referring to the experiments carried out at the Polytechnic
University of Turin, we have already stressed that \ the \ presence
of iron could be another factor, together with the energy release,
to be considered to account for the emission of neutrons from
granite and not from marble, and to open up prospects for future
experimental investigations into materials that fail in a
catastrophic mode. A subsequent investigation may be conducted both
on brittle iron-free materials and on ductile materials containing
iron, or on iron itself. Finally, after selecting the materials, it
would be necessary to analyse test specimens having different
dimensions and slenderness ratios, possibly by setting different
piston travel velocities on the testing machine. These studies would
contribute to an understanding of the influence of test specimens
dimensions and loading rate on neutron emission during fracture.

One can also think to carry out fatigue tests on the marble and
granite samples by increasing the pressure exerted on them up the
limit of fracture without reaching the fracturing of the material.
In this context, it will be possible to carry out cycles of these
compression tests and search for any emitted piezonuclear radiation
(neutrons, alpha, beta and gamma) during these cycles.

Instead of using ultrasounds or mechanical pressing machines, it is
possible to think of applying pressure on a pellet containing iron
by a process of ablation carried out by laser beams impinging on the
pellet.

A further method of applying pressure could be seen in the
experiments carried out by electrolytic reactors of the Mizuno type
\cite{miz}. The evidence of the existence of pressure waves in these
electrolytic reactors is supported by the presence of acoustic waves
that are continuously audible during the operation of these
equipments. It will be possible to carry out experiments by these
reactors and solutions of iron salts and verify whether piezonuclear
reactions are a consequence of variations of pressure in the
electrolytic plasma or whether there exists any contribution to
these reactions due to the local fracturing and morphological
changes of the electrodes by shock waves which are generated in the
plasma surrounding them.

The experiments carried out so far have been focused on the
detection and measurements of emitted neutrons and other possible
ionising radiation. Of course it will be possible to broaden our
view on these effects by carrying out other types of measurements.
Calorimetric measurements would be of great interest in order to
check the existence of extra heat and in order to have a further
experimental element to compare with the homologous obtained in CMNS
experiments \cite{storms}.

One further interesting perspective, both on the experimental side
and the theoretical one, is the chance to apply the concept of Local
Lorentz Invariance breakdown, which has been the starting point for
all of these experiments and considerations, to the issue of
sonoluminescence whose explanation is far from being clear.

\textbf{Acknowledgments}. We are greatly indebted to Fabio Pistella,
former President of the National Research Council of Italy, for his
invaluable encouragements, suggestions and remarks, concerning the
classical model of piezonuclear reactions, the detectors and hints
for improvements of the experiments. His deep knowledge of nuclear
reactions and devices was very precious in order to reach deeper
insights about the new physics of piezonuclear reactions. We are
also grateful to him for his sharp review of the manuscript and for
kindly helping us in writing some parts. We are obviously indebted
to all people who supported us in many ways in carrying out the
experiments: the military technicians of the Italian Armed Forces A.
Aracu, A. Bellitto, F. Contalbo, P. Muraglia; M. T. Topi, Director
of ARPA Laboratories of Viterbo; the following personnel of Casaccia
ENEA Laboratories: P. Giampietro, Director, G. Rosi, responsible of
the nuclear reactor "TAPIRO", and A. Santagata; G. Ingo and C.
Ricucci, of the Microscopy Laboratory ISMN-CNR and L. Petrilli, of
CNR-Rome 1 Area, Montelibretti, for performing the mass spectrometry
of the cavitated samples; R. Capotosto, for technical support on the
sonotrode tip; F. Rosetto; G. Cherubini and L. Stefani, of the
Laboratory cetli-nbc; A.Carpinteri, G.\ Lacidogna,and A. Manuello of
the Politechnic of Turin. On the theoretical side, invaluable
comments by E. Pessa are gratefully acknowledged. Thanks are also
due to F. Mazzuca, President of Ansaldo Nucleare, for deep interest
and warm encouragement.

\end{document}